\documentclass[a4paper,12pt]{article}
\usepackage{latexsym}
\usepackage{epsfig}
\usepackage[usenames]{color}

\def\){\right)}
\def\({\left( }
\def\]{\right] }
\def\[{\left[ }

\def\no{\nonumber \\}

\def\be{\begin{equation}}
\def\ee{\end{equation}}
\def\ba{\begin{eqnarray}}
\def\ea{\end{eqnarray}}
\def\no{\nonumber \\}

\def\ra{\rangle}
\def\la{\langle}

\def\a{\alpha}
\def \bi{\bibitem}

\def\b{\beta}

\def\F{\Phi}

\def\la{\langle}
\def\ra{\rangle}

\begin{document}
\begin{titlepage}
\begin{center}
 {\Large \bf Walls in supersymmetric massive \\
\vskip 0.2cm
 nonlinear sigma model on complex quadric surface}

\vskip 0.8cm
\normalsize
\renewcommand\thefootnote{\alph{footnote}}

\vskip 0.5cm

{\bf Masato Arai$^\dagger$\footnote{arai(at)sogang.ac.kr},
 Sunggeun Lee$^\ddagger$\footnote{sglkorea(at)hotmail.com}
 and Sunyoung Shin$^\sharp$\footnote{sihnsy(at)skku.edu}
}

\vskip 0.5cm

{\it $^\dagger {}^\sharp$ Center for Quantum Spacetime (CQUeST), Sogang University, \\
   Shinsu-dong 1, Mapo-gu, Seoul 121-742, Korea\\
   \vskip 0.5cm
 $^\ddagger$ Department of Physics and Research Institute for Basic Sciences, \\
   Kyung Hee University, Seoul 130-701, Korea \\
   \vskip 0.5cm
 $^\sharp$ Department of Physics, Sungkyunkwan University, \\
   Chunchun-dong 300, Jangan-gu, Suwon 440-746, Korea
}

\vskip 2cm

\begin{abstract}
The Bogomol'nyi-Prasad-Sommerfield (BPS) multiwall solutions
 are constructed in a massive K\"ahler nonlinear sigma model
 on the complex quadric surface, $Q^N={SO(N+2) \over SO(N)\times SO(2)}$
 in 3-dimensional space-time.
The theory has a nontrivial scalar potential
 generated by the Scherk-Schwarz
dimensional reduction from the massless nonlinear sigma
 model on $Q^N$ in 4-dimensional space-time and it gives rise to
 $2[N/2+1]$ discrete vacua.
The BPS wall solutions connecting these vacua are obtained based on
 the moduli matrix approach.
It is also shown that the moduli space of the BPS wall
solutions
 is the complex quadric surface $Q^N$.
\end{abstract}
\end{center}
\end{titlepage}

\renewcommand{\theequation}{\thesection.\arabic{equation}}
\newpage
\section{Introduction}
\setcounter{equation}{0}
It is well known that topological solitons play an important role in various fields
 in physics such as string theory, field theory,
 cosmology and condensed matter physics.
For the investigation of topological solitons
 supersymmetric (SUSY) field theories provide
 a nice arena since partial preservation of SUSY automatically gives
 the solution of equation of motion \cite{WittenOlive}.
This solution is called the Bogomol'nyi-Prasad-Sommerfield (BPS)
 state \cite{BPS}. One of the simplest BPS states is the so-called
 domain wall \cite{CQR,AT}, which
 is an extended object with codimension one.
Since it preserves half of the original SUSY, it is called a
 half BPS state. Such a solution has been well studied in various
 SUSY models.

In particular, recently there has been progress in constructing wall
 solutions in SUSY gauge theories with eight
 supercharges in four and five dimensions \cite{INOS1,INOS2}.
\footnote{By using the moduli matrix approach, various kinds of
 interesting solutions such as
 monopole-vortex-wall systems \cite{INOS3}, domain wall webs \cite{EINOS1},
 non-Abelian vortices \cite{EINOS2}, instanton-vortex systems \cite{EINOS3} and
 Skyrmions \cite{ENOT} were also found in $U(N_C)$ gauge theories.
 For a comprehensive review, see \cite{INOS2}.} In
\cite{INOS1,INOS2} SUSY $U(N_C)$ gauge theory coupled to
 $N_F(>N_C)$ massive flavors with
 the Fayet-Iliopoulos term has been considered and a systematic way to construct
 possible BPS domain walls has been formulated.
This formulation is called the moduli matrix approach. The mass term
gives rise to a nontrivial scalar potential, yielding
 ${}_{N_F} C_{N_C}$ number of discrete vacua.
Exact BPS multiwall solutions which interpolate these vacua
 with generic parameters covering the complete moduli space is
 obtained by taking the infinite gauge coupling.
For certain values for the finite gauge coupling limit, exact BPS
 multiwall solutions are also obtained.
It has been shown that the total moduli space of the BPS wall solutions
 is the Grassmann manifold,
  $G_{N_F,N_C}\equiv {U(N_F) \over U(N_C) \times U(N_F-N_C)}$.
The infinite gauge coupling limit yields vanishing kinetic
 terms of gauge fields and their superpartners.
These fields just become
 Lagrange multipliers giving constraints to matter fields.
In other words, the model becomes a quotient action of the massive
 hyper-K\"ahler (HK) nonlinear sigma model (NLSM) whose target metric
 is the cotangent bundle over the Grassmannian,
 $T^*G_{N_F,N_C}$.
This model was originally studied in \cite{ANS}.
The same number of discrete
 vacua was obtained there but the BPS wall solutions were not.
They were only known in massive HK NLSMs in the subclass of $T^*G_{N_F,N_C}$,
 especially, for $T^*G_{2,1}\simeq T^*{\bf C}P^1$ \cite{AT, GTT, GPTT, ANNS, AIN}
 until \cite{INOS1} appeared.

The Grassmann manifold is one of the compact Hermitian symmetric
 spaces (HSS).
The compact HSS consists of the four classical types
\be
 G_{N+M,M}= \frac{U(N+M)}{U(N)\times U(M)},~~\frac{SO(2N)}{U(N)},~~
 \frac{Sp(N)}{U(N)},~~Q^N=\frac{SO(N+2)}{SO(N)\times SO(2)},
\ee
and the two exceptional types
\be
 \frac{E_6}{SO(10)\times U(1)} ,~~\frac{E_7}{E_6 \times U(1)}.
\ee
It would be interesting to investigate domain walls in massive HK NLSMs on
 cotangent bundles over the HSSs other than the Grassmann manifold.
It is expected that they also possess
 discrete vacua and various kinds of domain walls connecting them.
The moduli matrix approach would help to construct domain wall solutions
 in these models.
In order to apply this approach to the above models, they have to be
 described as a quotient action, namely, SUSY gauge theories with infinite
 gauge coupling limit.
Massless HK NLSMs on the cotangent bundles over the classical HSSs
 \cite{KG1, KG2, AKL1}\footnote{A massless HK
 NLSM on the tangent bundle over the complex quadric surface being one of the classical HSSs has
 been worked out in \cite{AN}.} and over the $\frac{E_6}{SO(10)\times U(1)}$
 \cite{AKL2} were obtained in projective
 superspace \cite{KLR,LR}, but without using a gauge field Lagrange multiplier.
Actually, it is difficult to construct them as a quotient action.

On the other hand, it was observed that
 when considering vacua and domain walls in the massive HK NLSM on $T^*G_{N_F,N_C}$,
 the cotangent part is irrelevant \cite{ANNS, INOS1}.
In other words, in order to investigate them in this model,
 we can simply set the cotangent part to be zero.
In this setting, the massive HK NLSM on $T^*G_{N_F,N_C}$ reduces to the massive
 K\"ahler NLSMs on $G_{N_F,N_C}$.
The same situation would happen when considering massive HK NLSMs on cotangent
 bundles over HSSs other than $G_{N_F,N_C}$.
\footnote{For the case of the HSS ${\cal M}$, it is expected that the moduli
 space of domain walls is the base manifold ${\cal M}$ as in the case of the $T^* G_{N,M}$ model.
However, the moduli space of walls is not ${\cal M}$ in general.
Such an example has been examined in the NLSM on $T^*{\cal M}$ where
${\cal M}$
 is a special Lagrangian submanifold \cite{EINOOST}.
}

Inspired by this observation, in this paper,
 we study a massive K\"ahler NLSM on the complex quadric
 surface,
 $Q^N= {SO(N+2) \over SO(N)\times SO(2)}$.
We start with
 the massless K\"ahler NLSM on $Q^N$ in 4-dimensional space-time
 which has been formulated as a SUSY gauge theory in \cite{HN}.
Its massive version can be
 constructed via the
 Scherk-Schwarz dimensional reduction \cite{SS}.
Mass terms are characterized by the Cartan matrix of the isometry of
 the model, $SO(N+2)$ and give rise to a nontrivial scalar potential.
From the vacuum condition we
 find that the theory has $2\left[{N \over 2}+1 \right]$ discrete vacua.
We also find
 the exact domain wall solutions interpolating these vacua
 and the moduli spaces of the solutions.
The latter is shown to be the complex quadric surface.

Organization of this paper is as follows. In Section 2, we introduce
our model and investigate vacuum structure. We also derive
 half BPS equations.
In Section 3, exact solutions of the BPS equations are
 obtained in the use of the moduli matrix approach.
Section 4 is devoted to conclusion and discussion. In Appendix A, we
list moduli matrices of  multiwalls including compressed
 walls for the $N=4$ case.
In Appendix B, possible parameter regions for a quadruple wall in
 the $N=4$ case are given.

\section{Massive K\"ahler NLSM on $Q^N$} \label{sec;massiveKahler}
\setcounter{equation}{0} We start with a brief review of
 the massless NLSM on $Q^N$ in 4-dimensional space-time.
We basically follow the notation of \cite{WB}.

The SUSY gauge theory realizing the NLSM on $Q^N$ in terms of the ${\cal N}=1$ superfields
 in 4-dimensional space-time is given in \cite{HN}.
Let $\phi^i(x,\theta,\bar \theta)$ ($i=1,\cdots,N+2$) be chiral superfields,
 $\bar D_{\dot \alpha} \phi^i = 0$ belonging to a vector representation of $SO(N+2)$.
Introducing an auxiliary vector superfield $V (x,\theta,\bar \theta) \; (= V^\dagger)$
 and an auxiliary chiral superfield  $\phi_0(x,\theta,\bar \theta)$,
 being a singlet representation of $SO(N+2)$, the Lagrangian is described as
\begin{eqnarray}
 {\cal L} = \int d^4 \theta
 (\bar{\phi}^{i} \phi^i e^V - r^2 V)
 + \left( \int d^2 \theta \, \phi_0 (\phi^i)^2 + {\rm c.c.}\right),
 \label{Q^n-lag}
\end{eqnarray}
where $r^2$ is a real positive constant called the
Fayet-Iliopoulos parameter. Repeated indices $i$ are summed over
here. In the following this rule is implicitly assumed unless
stated. This Lagrangian possesses gauge invariance
\begin{eqnarray}
 V \to V - \Lambda - \Lambda^\dagger , \quad
 \phi^i \to e^{\Lambda} \phi^i, \quad
 \phi_0 \to e^{-2 \Lambda} \phi_0,
\end{eqnarray}
with an arbitrary chiral superfield $\Lambda(x,\theta,\bar \theta)$.
The equation of motion for $V$ is given by $\bar{\phi}^i\phi^i e^V -r^2=0$,
 which can be solved as $V = - \log (\bar{\phi}^{i}\phi^i/r^2)$.
If the superpotential is absent in the Lagrangian (\ref{Q^n-lag}),
 we obtain the K\"ahler potential of ${\bf C}P^{N+1}$.
In this case, substituting the solution back into the K\"ahler potential of
 (\ref{Q^n-lag}), we have
\begin{eqnarray}
  K = r^2 \log \left(1 + {\bar{\Phi}^a \Phi^a \over r^2}\right),
\end{eqnarray}
 with a gauge fixing $\phi^{\rm T} = (\Phi^{a},r)$($a=1,\cdots,N+1$).
The superpotential in (\ref{Q^n-lag}) gives an additional constraint
 through the equation of motion for $\phi_0$:
\ba
 (\phi^i)^2 = 0. \label{const-QS}
\ea
Therefore, the complex quadric surface is defined as a hypersurface embedded into
 the complex projective plane ${\bf C}P^{N+1}$ \cite{CV, KN}.
Let us solve the constraint (\ref{const-QS}). First, we decompose
$\phi^i$ into a representation of a $SO(N)\times U(1)$
 group of $SO(N+2)$ as $\phi^{\rm T}=(x,y^{I},z)$ where $x$ and $z$ are complex
 scalars, and $y^I(I=1,\cdots, N)$ is a complex vector.
Performing the unitary transformation \cite{HN}
\ba
 \phi \rightarrow
 \left(
  \begin{array}{ccc}
   {i \over \sqrt{2}} & 0 & {1 \over \sqrt{2}} \\
    0                 & {\bf 1}_{\bf N} & 0 \\
   -{i \over \sqrt{2}} & 0 & {1 \over \sqrt{2}}
  \end{array}
 \right)\phi,
\ea
then (\ref{const-QS}) becomes
\ba
  (\phi^i)^2 \rightarrow \phi^T J \phi = 2xz+(y^I)^2=0. \label{const-QS2}
\ea
Here $J$ is the rank 2 invariant tensor defined as
\begin{eqnarray}
J = \left(
    \begin{array}{ccc}
        0 &  {\bf 0}      &1 \\
 {\bf  0} & {\bf 1}_{n} &{\bf 0} \\
        1 &  {\bf 0}      &0
    \end{array}
            \right) .
\end{eqnarray}
The constraint (\ref{const-QS2}) can be solved to give
$
 \phi^{\rm T}=(x,y^I,-{(y^I)^2 \over 2x}).
$
Eliminating $V$ and making a gauge fixing as $\phi^T=(r,\F^I,-{(\F^I)^2 \over 2})$,
 we obtain the K\"ahler potential of the quadric surface \cite{Hua1, Hua2, MPS}
\begin{eqnarray}
  K = r^2 \log \left(1 + {\bar{\F}^I \F^I \over r^2}+{(\F^I)^2(\bar{\F}^I)^2 \over 4r^2}\right).
\end{eqnarray}

Next we derive a massive NLSM on $Q^N$.
In the above, starting from the quotient action, we eliminate the vector
 superfield $V$ and make a gauge fixing to obtain the known K\"ahler potential.
In order to utilize the formulation in \cite{INOS1, INOS2}
 we leave the vector superfield in the action as
 an independent degree of freedom.
Since we are interested in a solitonic solution, we focus only on
 the bosonic part of the Lagrangian in the following.
Substituting the expressions
\ba
 &&\phi^i(x,\theta,\bar{\theta}) = \phi^i(x) + \theta^2 F^i,\no
 &&\phi_0(x,\theta,\bar{\theta}) = \phi_0(x) + \theta^2 F_0,\no
 &&V(x,\theta,\bar{\theta}) = 2\theta \sigma^\mu \bar{\theta} v_\mu +{1 \over 2}\theta^2\bar{\theta}^2 D,
\ea
into (\ref{Q^n-lag}), the bosonic part of the Lagrangian becomes (we take $r=1$ for simplicity)
\ba
{\cal L}_{\rm bos}&=&-\partial_\mu \phi^i \partial^\mu \bar{\phi}^{i} +|F^i|^2
 -i v_\mu (\bar{\phi}^{i} \partial^\mu \phi^i-\partial^\mu \bar{\phi}^i \phi^i)
 -v^\mu v_\mu \bar{\phi}^{i} \phi^i +{1\over 2}D(\bar{\phi}^{i}\phi^i-1)\no
 && +F_0(\phi^i)^2+\bar F_0(\bar\phi^i)^2 + 2\phi_0\phi^iF^i + 2\bar\phi_0\bar\phi^i\bar F^i. \label{comp-q}
\ea
The Greek letter $\mu$ denotes a 4-dimensional space-time index.
Eliminating the auxiliary fields, we obtain the scalar potential

\be
 V= |F^i|^2 =4 |\phi_0|^2 |\phi^i|^2,
\ee
and the constraints
\be
 (\phi^i)^2 = 0,\,\,\,\,\,\,(\bar{\phi}^i)^2=0,\,\,\,\,\,\,|\phi^i|^2-1=0. \label{const-Q}
\ee
The vacuum condition $V=0$ tells us that $\phi_0=0$ or $\phi^i=0$.
The latter is inconsistent with the last condition in
(\ref{const-Q})
 while the former solution is consistent and leads to $\phi^i \neq 0$.
However, the former one does not give discrete vacua. Therefore, no
 domain wall solution exists in this case.

The situation changes when
 mass terms are introduced in the above model.
We perform the Scherk-Schwarz dimensional reduction
 for the generation of mass \cite{SS}.
Specifying that fields in the
 $x^3$ direction move along orbits of the Killing
 vectors $f(\phi)$ and $\bar{f}(\bar{\phi})$ in the quadric
 surface
\be
 {\partial \phi^i \over \partial x^3}=f^i(\phi) = -i M^{ij}\phi^j,\,\,\,\,
 {\partial \bar{\phi}^i \over \partial x^3}=\bar{f}^i(\bar{\phi}) =  i (M^{ij}\phi^j)^\dagger, \label{SS}
\ee
where $M^{ij}$ is the Cartan matrices of $SO(N+2)$ given by
\begin{eqnarray}
 && M^{ij} = \sum_{a=1}^{[{N \over 2}+1]}m_a \delta_{aa}
 \otimes \sigma_2\,\,\,\,\,\,(\mbox{even}\,\,N), \\
&& M^{ij} = \left(
 \begin{array}{cc}
 \displaystyle \sum_{a=1}^{[N/2+1]}m_a \delta_{aa}\otimes \sigma_2 & 0 \\
   0 & 0
 \end{array}
 \right) \,\,\,\,\,\,(\mbox{odd}\,\,N).
\end{eqnarray}
Here $m_a$ is a real
 mass parameter and $\delta_{aa}$ is the $[N/2+1] \times [N/2+1]$ unit matrix.
The matrix $M^{ij}$ can be generic if we take $m_a\neq 0$ for every $a$ and $m_a^2 \neq m_b^2$ for $a\neq b$.
In this paper, we further assume that $m_a >m_{a+1}>0$.
Note that by introducing mass terms, flavor symmetry $SO(N+2)$ is broken down to $SO(2)^{[N/2+1]}$.

Substituting (\ref{SS}) into the component action (\ref{comp-q}), we have
\ba
{\cal L}_{\rm bos}&=& - \partial^m \bar{\phi}^{i}\partial_m \phi^i  -|f^i|^2 + |F^i|^2
    -iv_m (\bar{\phi}^{i} \partial^m \phi^i-\partial^m \bar{\phi}^i \phi^i)
    -i \sigma (\bar{\phi}^{i} f^i(\phi)-\bar{f}^i(\bar{\phi}) \phi^i) \no
 && -(v^m v_m+\sigma^2) \bar{\phi}^{i} \phi^i +{1\over 2}D(\bar{\phi}^{i}\phi^i-1) \no
 && +F_0(\phi^i)^2+\bar F_0(\bar\phi^i)^2 + 2\phi_0\phi^iF^i + 2\bar\phi_0\bar\phi^i\bar F^i,
\ea
where $\sigma=v_3$.
A Roman letter index $m$ refers to the first three components of the
 4-dimensional index $\mu$.
Eliminating the auxiliary fields $F^i, F_0$ and $D$, we have
\ba
{\cal L}_{bos}&=&-\partial_m \phi^i \partial^m \bar{\phi}^{i}
    -i v_m (\bar{\phi}^{i} \partial^m \phi^i-\partial^m \bar{\phi}^i \phi^i)
    -v^m v_m \bar{\phi}^{i} \phi^i-V,
\ea
with the constraints (\ref{const-Q}).
The scalar potential $V$ is given
by
\ba
V = \left|f^i - i\sigma \phi^i \right|^2 + 4|\phi_0|^2|\phi^i|^2. \label{pot-SS}
\ea
The first term comes from the dimensional reduction and it gives
 rise to discrete vacua as we will see below. The vacuum condition is
readily read off as
\ba
 f^i(\phi) - i\sigma \phi^i = 0, \label{cond1}
\ea
and
\ba
 \phi_0 =0\quad\mbox{or}\quad \phi^i=0, \label{cond2}
\ea
with the constraints (\ref{const-Q}).
The latter solution in (\ref{cond2}) is inconsistent with the last constraints
 in (\ref{const-Q}).
Therefore, we shall consider the case, $\phi_0=0$ and $\phi^i\neq 0$.
The condition (\ref{cond1}) is rewritten by
\ba
0 = \displaystyle
    |f^i(\phi) - i\sigma \phi^i|^2 =
    \displaystyle
    \left(\bar{\phi}^{2a-1},\bar{\phi}^{2a} \right)
    \left(
     \begin{array}{cc}
      \sigma & i m_a \\
      -i m_a & \sigma
     \end{array}
    \right)^2
    \left(
     \begin{array}{c}
      \phi^{2a-1} \\
      \phi^{2a}
     \end{array}
    \right)+ c|\sigma \phi^{N+2}|^2,
\label{cond1-1}
\ea
where $a$ is the flavor index running from $1$ to $[N/2+1]$
 and $c$ takes $0$ for even $N$ cases and $1$ for odd $N$ cases.
For later convenience, we perform the unitary transformation (it makes the Bogomol'nyi
 completion of the Hamiltonian easy as will be seen in the next section)
\ba
   \left(
    \begin{array}{c}
     \phi^{2a-1} \\
     \phi^{2a}
    \end{array}
   \right) \rightarrow
   \Phi^{\alpha a}\equiv
   \left(
    \begin{array}{c}
     \Phi^{1a} \\
     \Phi^{2a}
    \end{array}
   \right)
 = {1 \over \sqrt{2}}
   \left(
    \begin{array}{cc}
     1 & -i \\
     1 & i
    \end{array}
   \right)
   \left(
    \begin{array}{c}
     \phi^{2a-1} \\
     \phi^{2a}
    \end{array}
   \right), \label{rot}
\ea
where $\a=1,2$. Equation (\ref{cond1-1}) is rewritten by
\ba
 0=\sum_{i=1}^{\left[{N \over 2} + 1 \right]} \sum_{\alpha=1}^2
   |\lambda_{\alpha a} \Phi^{\alpha a}|^2
     +c|\sigma \phi ^{N+2}|^2, \quad \lambda_{\a a}\in {\bf R}, \label{cond1-2}
\ea
where $\lambda_{1a}=\sigma + m_a$ and $\lambda_{2a}=\sigma - m_a$.
The constraints (\ref{const-Q}) become
\ba
&& |\Phi^{\alpha a}|^2 + c |\phi^{N+2}|^2 = 1, \label{const-Q1} \\
&&  2\Phi^{1 a} \Phi^{2 a} + c(\phi^{N+2})^2 = 0
    \quad \mbox{and~~c.c.} \label{const-Q2}
\ea
In the following, we solve the set of these equations
 for even and odd $N$ cases, separately.

\noindent{\bf a)Even $N$ case}\\
In this case, Equation (\ref{cond1-2}) tells us that
\ba
 \lambda_{\alpha a} \Phi^{\alpha a} = 0.\quad \quad (\mbox{no~sum~for~}\a, a) \label{cond1a}
\ea
Equation (\ref{cond1a}) leads to $\lambda_{\alpha a}=0$  or
 $\Phi^{\alpha a}=0$. Among them the former one is only consistent
 with (\ref{const-Q1}). It gives two solutions $\sigma=-m_a$ and
 $\sigma=m_a$. The first case says that $\Phi^{1a}\neq 0$ and
 $\Phi^{2a}=0$ for some $a$. The constraint (\ref{const-Q2}) is
 satisfied with this solution while (\ref{const-Q1})
 gives $|\F^{\a a}|^2=1$.
Therefore, a solution in this case is
\ba
 \Phi^{\a a}=\left(
  \begin{array}{ccccccc}
   0 & \cdots & 0 & 1 & 0 & \cdots & 0 \\
   0 & \cdots & 0 & 0 & 0 & \cdots & 0
  \end{array}
 \right),\quad \sigma=-m_a, \label{sol1}
\ea
where a phase is set to be zero by the flavor symmetry $SO(2)^{[N/2+1]}$.
There exit $[N/2+1]$ solutions as the index runs from 1 to $[N/2+1]$.
Similarly we can analyze for the second case $\sigma=m_a$. A
solution for some $a$ is given by
\ba
 \Phi^{\a a}=\left(
  \begin{array}{ccccccc}
   0 & \cdots & 0 & 0 & 0 & \cdots & 0 \\
   0 & \cdots & 0 & 1 & 0 & \cdots & 0
  \end{array}
 \right),\quad \sigma=m_a. \label{sol2}
\ea
Again we see that there are $[N/2+1]$ solutions.
Taking into account both cases, we find that the theory has $2[N/2+1]$ vacuum solutions.

\noindent{\bf b) Odd $N$ case}\\
In this case, the vacuum condition is given by
\ba
\sigma \phi^{N+2}=0, \label{cond1b}
\ea
in addition to (\ref{cond1a}).
From (\ref{cond1b}), we have $\sigma=0$ or $\phi^{N+2}=0$.
The former solution with (\ref{cond1a}) leads to
 $\Phi^{\alpha a}=0$ and $|\phi^{N+2}|^2 = 1$.
This is inconsistent with (\ref{const-Q2}). It is therefore not a
 solution. Considering the $\phi^{N+2}=0$ case, the situation turns
 out to be the same as the even $N$ case. Therefore, we find that there exist $2[N/2+1]$
 vacuum solutions for the odd $N$ case.

We make comments in order. For the $N=1$ case, the target metric of
 our model is $Q^1$
 which is isomorphic to ${\bf C}P^1$ and the theory has two discrete vacua.
This number of vacua
 is the same with one in the massive $T^*{\bf C}P^1$ NLSM model \cite{ANNS}.
As mentioned in the Introduction,
 the cotangent part of the massive $T^*{\bf C}P^1$ model is irrelevant
 when considering vacua and wall solutions.
Therefore, we find that our vacuum solution for the $N=1$ case is
 consistent with the result of the massive $T^*{\bf C}P^1$ model. For the $N=4$ case,
 the target space becomes $Q^4$ which is isomorphic to
 Grassmannian, $G_{4,2}$. In this case, there exist six vacua. On the other hand,
 it is known that there are ${}_{N_F} C_{N_C}$ vacuum solutions in the massive
 NLSM on $T^*G_{N_F,N_C}$ \cite{ANS},
 yielding six vacua for the $T^*G_{4,2}$ case. Repeating the same discussion
 as in the $N=1$ case we again find that our result is
 consistent.
New results appear in other cases.
For instance,
 for the $N=2$ and $N=3$ cases, target spaces of our model are isomorphic to
 ${\bf C}P^1\times {\bf C}P^1$ and $Sp(2)/U(2)$, respectively.
 There exist four vacua in both cases.
For $N>4$, there is no isomorphism
 and this is therefore purely the result
 of the complex quadric surface.

%
%
\section{BPS equations}
\setcounter{equation}{0}
In this section, we derive the BPS equation through the Bogomol'nyi completion of
 (the bosonic part of) the Hamiltonian.
Since we are interested in a time-independent wall solution,
 we assume that fields have no time dependence,
 $\partial_0 \phi^i=0$ and
 that all fields depend on the coordinate of only one dimension
 of $x_1$, which we shall write $x$.
We also assume
 the Poincar\'e invariance on the two-dimensional world volume of the wall,
 which implies $v_0=v_2=0$.
The energy along the $x$ direction is given by
\ba
E&=&\int d x (|D_1 \phi^i|^2+|f^i - i\sigma \phi^i|^2 + 4|\sigma \phi_0|^2) \no
 &=& \int d x {\Bigg \{}
     \sum^{[{N\over 2}+1]}_{a=1}\sum^2_{\alpha=1}
     \left(|D_1 \Phi^{\alpha a}|^2
     + |\lambda_{\alpha a} \Phi^{\alpha a}|^2 + 4 |\phi_0 \Phi^{\alpha a}|^2 \right) \no
 & &\quad \quad + c(|D_1 \phi^{N+2}|^2 + \sigma^2 |\phi^{N+2}|^2+ 4|\phi_0 \phi^{N+2}|^2) {\Bigg \}},
\ea
with the constraints (\ref{const-Q1}) and (\ref{const-Q2}).
The covariant derivative is defined by $D_1 \Phi^{\alpha a} = (\partial_1 - iv_1)
 \Phi^{\alpha a}$.
The Bogomol'nyi completion of the energy can be performed as
\begin{eqnarray}
E &=& \int d x {\Bigg\{}
      \sum^{[{N\over 2}+1]}_{a=1} \sum^2_{\alpha=1}
      \left(|D_1 \Phi^{\alpha a} \mp \lambda_{\alpha a} \Phi^{\alpha a}|^2
      + 4 |\phi_0 \Phi^{\a a}|^2 \right) \no
  & & + c \left(|D_1 \phi^{N+2} \mp \sigma \phi^{N+2}|^2
      + 4 |\phi_0 \phi^{N+2}|^2 \right)\pm T{\Bigg\}} \ge \pm T, \label{h}
\end{eqnarray}
where $T$ is a tension defined by
\ba
T \equiv \int d x \sum^{[{N\over 2}+1]}_{a=1}
            \partial_1  m_a (|\Phi^{1a}|^2-|\Phi^{2a}|^2).
\ea
From (\ref{h}) the BPS equations are obtained as
\ba
 && D_1 \Phi^{\alpha a} \mp \lambda_{\alpha a} \Phi^{\alpha a}=0, \quad
    (\mbox{no~sum~for~}\a,~a)\label{bps1} \\
 && \phi_0 \Phi^{\alpha a}=0, \label{bps2}\\
 && D_1 \phi^{N+2} \mp \sigma \phi^{N+2}=0, \label{bps3} \\
 && \phi_0 \phi^{N+2}=0.\label{bps4}
\ea
Equations (\ref{bps2}) and (\ref{bps4}) tell us that $\phi_0=0$ or
 $\Phi^{\alpha a}=\phi^{N+2}=0$, but the latter solution is inconsistent
 with (\ref{const-Q1}).
Taking the former solution, the BPS equations are simplified to be
 (here we take upper sign in (\ref{bps1}) and (\ref{bps3}))
\ba
 && D_1 \Phi^{\alpha a} - \lambda_{\alpha a} \Phi^{\alpha a} = 0,
    \quad (\mbox{no~sum~for~}\a,~a)\label{bps5} \\
 && D_1 \phi^{N+2} - \sigma \phi^{N+2}=0. \label{bps6}
\ea

%
%
\section{BPS wall solution}
\subsection{BPS equations}
\setcounter{equation}{0}
In this section we solve the BPS equations (\ref{bps5}) and (\ref{bps6})
 together with the constraints (\ref{const-Q1}) and (\ref{const-Q2}) by using
 the moduli matrix approach \cite{INOS1,INOS2}.
First of all, we introduce a complex function $S(x)$ defined by
\ba
 -\sigma - i v_1 = S^{-1}(x)\partial_1 S(x). \label{b1}
\ea
Let us change variables from $\Phi^{\a a}$ and $\phi^{N+2}$ to complex valued functions
 $f^{\a a}$ and $f^{N+1}$ by using $S$
\ba
 \Phi^{\a a} \equiv S^{-1}f^{\a a},\quad \phi^{N+1}\equiv S^{-1}f^{N+2}.\label{b2}
\ea
Substituting (\ref{b1}) and (\ref{b2}) into (\ref{bps5}) and (\ref{bps6}), we have
\ba
 \partial_1 f^{\a a} = (\hat{M}_a)^\a_{~\b} f^{\b a},\quad \partial f^{N+2}=0, \quad (\mbox{no~sum~for~} a)
 \ea
where $\hat{M}_a\equiv {\rm diag}(m_a, -m_a)$.
It can be easily solved as
\ba
 f^{\a a} = (e^{\hat{M}_a x})^\a_{~\b}H_0^{\b a},\quad f^{N+2}=H_0^{N+2},  \quad (\mbox{no~sum~for~} a)
 \label{b3}
\ea
with a complex constant matrix $H_0^{\a a}$ and a complex constant $H_0^{N+1}$
 as integration constants.
Since they include information of vacua and positions of walls, it is called the moduli
 matrix \cite{INOS1,INOS2}.
From (\ref{b3}), $\Phi^{\a a}$ and $\phi^{N+2}$ can be solved in terms of $S$ as
\ba
 \Phi^{\a a} = S^{-1} (e^{\hat{M}_a x})^\a_{~\b} H_0^{\b a}, \quad
 \phi^{N+2}=S^{-1} H_0^{N+1}. \quad (\mbox{no~sum~for~} a)\label{b4}
\ea
The definitions (\ref{b1}) and (\ref{b2}) show that a set $(S,H_0)$ and another set
 $(S^\prime,H_0^\prime)$ give the same original fields $\sigma,~v_1,~\F^{\a a},$
 and $\phi^{N+2}$, provided that they are related by
\ba
 S^\prime = VS,\quad H_0^{\a a \prime} = V H_0^{\a a},\quad
 H_0^{N+1 \prime}=V H_0^{N+1}, \label{equiv}
\ea
where $V\in {\bf C}^*={\bf C}-\{0\}$.
This transformation $V$ defines an equivalent class among sets of the functions
 $(S,H_0^{\a a},H_0^{N+2})$ which represent physically equivalent results.
This kind of symmetry is called the world-volume symmetry \cite{INOS1}.
It is seen that the equivalence relation (\ref{equiv}) with the constraints
 (\ref{const-Q1}) and (\ref{const-Q2}) defines the complex quadric surface.
Making the unitary transformation,
\ba
   \left(
    \begin{array}{c}
     H_0^{1a} \\
     H_0^{2a}
    \end{array}
   \right)
   \rightarrow
   \left(
    \begin{array}{c}
     H_0^{2a-1} \\
     H_0^{2a}
    \end{array}
   \right)
 = {1 \over \sqrt{2}}
   \left(
    \begin{array}{cc}
     1 & 1 \\
     i & -i
    \end{array}
   \right)
   \left(
    \begin{array}{c}
     H_0^{1a} \\
     H_0^{2a}
    \end{array}
   \right),
\ea
and defining the vector $H_0^i \equiv (H_0^{2a-1},H_0^{2a},H_0^{N+2})$
 ($i=1,\cdots,N+2$), then Eqs. (\ref{equiv}), (\ref{const-Q1}) and (\ref{const-Q2}) are
\ba
 & S^\prime = VS,\quad H_0^{i \prime} = V H_0^{i}, \label{equiv2} & \\
 & \displaystyle
   |H_0^i|^2=1,\quad (H_0^i)^2=0,\quad (\bar{H}_0^i)^2=0.&
\ea
This is nothing but the definition of the complex quadric surface \cite{KN}.
Therefore, the moduli space of the domain walls is the complex quadric surface.

Since the BPS equation for matter parts are solved by means of the function $S$,
 the remaining task is to solve the constraints (\ref{const-Q1}) and (\ref{const-Q2}).
Substituting (\ref{b4}) into (\ref{const-Q1}) and (\ref{const-Q2}), we have
\ba
 & \displaystyle
    H_{0\a}^{\dagger~a} (e^{2\hat{M}_a x})^\a_{~\b} H_{0}^{\b a} + c |H_0^{N+2}|^2 = SS^\dagger, &
     \label{const-Q3} \\
 & \displaystyle
   2 H_0^{1 a} H_0^{2 a} + c(H_0^{N+2})^2 = 0 \quad \mbox{and~~c.c.}& \label{const-Q4}
\ea
Once a moduli matrix is given, $S$ is also obtained by (\ref{const-Q3}) and
 eventually the explicit solutions $\Phi^{\a a}$ and $\phi^{N+2}$ are
 obtained.
A form of moduli matrix should be determined so that it includes information of vacua,
 boundary conditions and positions of walls.
In addition, it must satisfy the constraint (\ref{const-Q4}).
In the next section, we show various types of possible moduli matrices and
 investigate their properties.

\vspace{5mm}
\subsection{Properties of moduli matrix} \label{sec;moduli matrix}
\noindent{\bf a)Vacuum}\\
This is the simplest example of the moduli matrix.
For the even $N$ case, the moduli matrices $H_0^{\a a}$ corresponding to vacua
 are given by
\ba
&& \hspace{38mm}{\mbox{\tiny k-th}} \nonumber \\
&& H_{0\la k \ra}=\left(
  \begin{array}{ccccccc}
   0 & \cdots & 0 & 1 & 0 & \cdots & 0 \\
   0 & \cdots & 0 & 0 & 0 & \cdots & 0
  \end{array}
 \right),\quad \sigma=-m_k, \label{sol1-H}
\ea
or
\ba
&& \hspace{38mm}{\mbox{\tiny k-th}} \nonumber \\
&& H_{0\la k \ra}=\left(
  \begin{array}{ccccccc}
    0 & \cdots & 0 & 0 & 0 & \cdots & 0 \\
    0 & \cdots & 0 & 1 & 0 & \cdots & 0
   \end{array}
 \right),\quad \sigma=m_k, \label{sol2-H}
\ea
where the index $k$ in $H_{0\la k \ra}$ labels the $k$-th vacua.
In what follows, we represent $\la k\ra$ as the $k$-th vacuum. These forms trivially satisfy (\ref{const-Q4}).
One can easily check that they yield the vacuum solutions (\ref{sol1})
 and (\ref{sol2}).
Substituting them into (\ref{const-Q3}), we have
\ba
 &&  S_{\la k \ra}=e^{m_k x} \quad \mbox{for~}\sigma=-m_k, \label{S-sol1}\\
 &&  S_{\la k \ra}=e^{-m_k x} \quad \mbox{for~}\sigma=m_k. \label{S-sol2}
\ea
Here we take the phase to be zero by using the flavor symmetry $SO(2)$. Substituting
 (\ref{S-sol1}) and (\ref{S-sol2}) with (\ref{sol1-H}) and (\ref{sol2-H}) into (\ref{b4}),
 the vacuum solutions (\ref{sol1}) and (\ref{sol2}) are obtained.

For the odd $N$ case, we just take into account $H_0^{N+2}=0$ in addition to
 (\ref{sol1-H}) and (\ref{sol2-H}).
Repeating the same analysis as was performed in the even $N$ case,
 one can see that they satisfy the constraint (\ref{const-Q4}) and give the correct vacua.

\noindent {\bf b)Single wall}\\
A simple example of a nontrivial configuration is a wall
configuration connecting two vacua,
 which we call a single wall.
First we consider the even $N$ case.
In this case, a moduli matrix $H_0^{\a a}$ representing a single wall
 connecting two vacua is written by two nonzero components.
For example, a moduli matrix satisfying (\ref{const-Q4}) is given by
\vspace{0.5cm} \ba &&\hspace{41.5mm}\mbox{\tiny
k-th}\hspace{5mm}\mbox{\tiny k+1-th} \no &&H_{0\la k \leftarrow k+1
\ra}=\left(
  \begin{array}{ccccccc}
  \cdots & 0 & e^{r_k} & e^{r_{k+1}} & 0 & \cdots \\
  \cdots & 0 & 0       & 0           & 0 & \cdots
   \end{array}
 \right),\quad r_i \in {\bf C}, \label{single1}
\ea
where we parametrize nonzero factor by exponent with complex
constants
 $r_i~(i=k,~k+1)$ for convenience\footnote{We follow the same parametrization as
 in \cite{INOS1}.}
 and take one exponential factor to be a unit by using the
 world-volume symmetry transformation (\ref{equiv}), namely, $r_k=0$.
Here the suffix $\langle k \leftarrow k+1 \rangle$ denotes the moduli matrix
 describing the BPS state interpolating from the vacuum $\la k+1 \ra$ at
 $x=-\infty$ to the vacuum $\la k \ra$ at $x=\infty$.

One can check that this moduli matrix gives
 the vacua at boundaries, $x=\pm \infty$.
In order to see that, notice that the solution for $\Phi^{\a a}$ and
 $\phi^{N+2}$ in (\ref{b4}) implies the transformation of the moduli matrix
\ba
 H_0^{\a a}\rightarrow (e^{\hat{M}_a x_0})^\a_{~\b}H_0^{\b a},\quad H_0^{N+2}\rightarrow H_0^{N+2}, \quad (\mbox{no~sum~for~} a) \label{shift}
\ea
under a translation $x \rightarrow x+x_0$.
For the case of (\ref{single1}), we have
\ba
 H_0^{1b}\rightarrow e^{m_b x_0}H_0^{1b},~~(b=k~\rm{or}~k+1)
 \label{shift1-H}
\ea
while $H_0^{2a}$ remains to be zero.
Since the world-volume symmetry transformation (\ref{equiv}) allows us to multiply $H_0^{1a}$ by
 the factor $V=e^{-m_k x_0-r_k}$, we have
\ba
 V H_0^{1a} = (\cdots, 0, 1, {\cal O}(e^{-(m_k - m_{k+1})x_0}),0,\cdots).
\ea
Taking $x_0\rightarrow \infty$, one sees that the moduli matrix becomes
 $H_{0\la k \ra}$ given by
(\ref{sol1-H}). Similarly, multiplying $H_0^{1a}$ by $V=e^{-m_{k+1} x_0-r_{k+1}}$ and taking
 $x_0\rightarrow -\infty$, one sees that
 the moduli matrix becomes $H_{0\la k+1 \ra}.$

The following moduli matrix is also possible to express a single wall
\ba
&&H_{0\la l-1 \leftarrow l \ra}=\left(
  \begin{array}{ccccccc}
  \cdots & 0 & 0        & 0           & 0 & \cdots \\
  \cdots & 0 & e^{r_{l}} & e^{r_{l-1}} & 0 & \cdots
   \end{array}
 \right).\label{ssingle2}\\
&&\hspace{41.5mm}\mbox{\tiny l-th}\hspace{5mm}\mbox{\tiny l-1-th}
\nonumber \ea
Repeating the same analysis as in the previous case, it can be seen that
 the moduli matrix gives
 the vacua $\la l \ra$ and $\la l-1 \ra$ at $x=-\infty$ and $x=\infty$, respectively.
In this case the most left (right) nonzero component represents the
vacuum
 at $x=-\infty~(x=\infty)$ unlike the previous case because opposite signs of
 masses appear in the shift of $H_0^{2a}$ according to (\ref{shift}).

Another possible choice for wall configurations is
 to take one nonzero component both in the first and the second lines
 in $H_0^{\a a}$, respectively.
For example,
\ba
&&H_{0\la k\leftarrow l\ra}=\left(
  \begin{array}{ccccccc}
  \cdots & 0 & e^{r_k} & 0       & 0 & \cdots \\
  \cdots & 0 & 0       & e^{r_l} & 0 & \cdots
   \end{array}
 \right),
\label{single3}
\ea
which also satisfies (\ref{const-Q4}).
In this case, one can check that the vacuum $\la k \ra$ is at $x=\infty$ and
 the vacuum $\la l \ra$ is at $x=-\infty$.

The following moduli matrix is not allowed as a single wall configuration for
 the even $N$ case
\ba
&&H_{0\la k\leftarrow l \ra}=\left(
  \begin{array}{cccccc}
  \cdots & 0 & e^{r_k} & 0 & \cdots \\
  \cdots & 0 & e^{r_l} & 0 & \cdots
   \end{array}
 \right),
\label{single4}
\ea
since it does not satisfy the constraint (\ref{const-Q4}).

Next we consider the odd $N$ case.
In this case, we just take into account $H_0^{N+2}$ in (\ref{const-Q3})
 and (\ref{const-Q4}).
It is easy to see that configurations such as (\ref{single1}),
 (\ref{ssingle2}) and (\ref{single3}) are possible with
 $H_{0}^{N+2}=0$.
In addition to these configurations,
 (\ref{single4}) is also allowed since
 (\ref{const-Q4}) can be satisfied if $H_0^{N+2}$ has the following
 values:
\ba
 H_0^{N+2}=\sqrt{2} i e^{(r_k+r_l)/2}. \label{odd-H}
\ea
The nonzero value of $H_0^{N+2}$ gives a nontrivial
 configuration of $\phi^{N+2}$
 through (\ref{b4}).
In this case, the moduli matrix (\ref{single4}) gives the vacuum $\la k \ra$ at $x=\infty$
 and the vacuum $\la l \ra$ at $x=-\infty$ while $H_0^{N+2}$ approaches $0$ at both boundaries.

From the above observation, we can see that the most left nonzero
component
 in the first line of $H_0^{\a a}$ represents a vacuum at $x=\infty$
 while a vacuum represented by the most
 left nonzero component in the second line is at $x=-\infty$.
It is also true for general (multiwall) cases:\vspace{0.5cm}
\ba
&&\hspace{38mm}\mbox{\tiny 1st} \hspace{10mm}\raisebox{-0.6em}{${\scriptstyle \longleftarrow}$}
  \hspace{-6mm}\raisebox{0.2em}{$\scriptstyle x\rightarrow \infty$} \no
&&H_{0\la 1\leftarrow k \ra}=\left(
  \begin{array}{cccccccc}
  \cdots & 0 &  e^{r_1} & * & \cdots  &   &\\
         & 0 &  \cdots  & 0 & e^{r_k} & * & \cdots
   \end{array}
 \right) \uparrow {\scriptstyle x \rightarrow \infty}, \\
&&\hspace{51mm}\mbox{\tiny k-th} \hspace{4mm}\raisebox{-0.6em}{${\scriptstyle \longrightarrow}$}
  \hspace{-6mm}\raisebox{0.2em}{$\scriptstyle x\rightarrow \infty$} \nonumber
\ea
where the asterisk expresses either zero or nonzero components. We
will show possible forms of multiwall configurations.

Before going to the discussion of multiwall configurations,
 we give some definitions concerning single walls.
Single wall configurations are classified into two types.
For even and odd $N$ cases of the moduli matrix given by
\ba
&&H_{0\la k\leftarrow l \ra}=\left(
  \begin{array}{cccccccc}
  \cdots & 0 & e^{r_k} & \underbrace{0  \cdots  0}_n & e^{r_{l}} & 0 & \cdots \\
  \cdots & 0 & 0  &              \cdots  & 0 & 0  & \cdots&
   \end{array}
 \right), \label{comp1}
\ea
this state defines an elementary wall or a compressed wall if $n=0$ or $n\neq 0$, respectively \cite{INOS1}.
Here $n$ is called the level of the single wall.
They appear in a different form of the moduli matrix which has
 a nonzero component in the first and the second lines in
 $H_0^{\a a}$, respectively,
\ba
&&H_{0\la k\leftarrow l \ra}=\left(
  \begin{array}{cccccccc}
  \cdots & 0 & e^{r_k}\hspace{-1.8mm} & \,\,\underbrace{0 \cdots \cdots\, \cdots \,\cdots 0}_n  \\
  \cdots & 0 & \cdots  &~0~~e^{r_l}~~\underbrace{0 \cdots \cdots 0}_m
   \end{array}
 \right). \label{comp2}
\ea
For the even (odd) $N$ cases,
 the configuration represents an elementary wall or a compressed wall if $n+m=1(n+m=0)$
 or $n+m>1(n+m>0)$, respectively.
Compressed walls are obtained as compression of a multiwall
configuration. We will see in detail through the explicit examples
 which will be explained in the next section.

%
%
\noindent{\bf c)Multiwalls}\\
It is easy to extend above configurations into multiwalls interpolating discrete vacua.
Let us consider the even $N$ case first.
A simple configuration
 connecting $n$ vacua is given by, for example,
\ba
H_{0\la 1\leftarrow n\ra}=\left(
  \begin{array}{cccccccc}
  \cdots & 0 & e^{r_1} & e^{r_2} & \cdots  & e^{r_{n}} & 0 & \cdots \\
  \cdots & 0 & 0       & 0       & \cdots  & 0         & 0 &\cdots
   \end{array}
 \right),\quad n \le N/2+1, \label{up}
\ea
which trivially satisfies the constraint (\ref{const-Q4}).
In what follows we will show that this configuration interpolates multiple vacua.
With the use of (\ref{shift}),
 $H_0^{1a}$ transforms as
\ba
 H_0^{1a}\rightarrow e^{m_a x_0}H_0^{1a},~~(\mbox{no~sum~for~}a) \label{shift-H-m}
\ea
while $H_0^{2a}$ remains to be zero.
Multiplying $H_0^{1a}$ by the factor
 $V=e^{-m_l x_0-r_l}$, the vector $VH_0^{1a}e^{m_a x_0}$ becomes
\ba
 VH_0^{1a}e^{m_a x_0}&=&(\cdots, e^{(m_{l-1}-m_l)(x_0-X_{l-1})},1,e^{-(m_{l}-m_{l-1})(x_0-X_{l})}, \cdots), \label{vac1}
\ea
where we have defined
\ba
X_l \equiv -{r_l-r_{l+1} \over m_l - m_{l+1}},\quad l=1,\cdots, n. \label{position}
\ea
We denote ${\rm Re}(X_l) = x_l$.
If we assume
\ba
 x_1 \gg x_2 \gg \cdots \gg x_{n}, \label{region}
\ea
 and consider the region of
 $x_{l-1} \gg x_0 \gg x_{l}$, then we see that in (\ref{vac1}) the $l$-th flavor component
 becomes dominant while the other components become negligible:
\ba
 e^{-m_l x_0-r_l}H_0^{1k}e^{m_k x_0}\sim \delta^{lk}.
\ea
By this way, we can specify the $l$-th vacuum.
Since $l$ runs from $1$ to $n$ in this case, it is found that
 the moduli matrix (\ref{up}) realizes $n$ number of discrete vacua.
As $x_0$ decreases (increases), the dominant element shifts to the
 right (left) gradually
 in the flavor space as $\delta^{lk}\rightarrow \delta^{(l-1)k}(\delta^{lk}\rightarrow \delta^{(l+1)k})$.
This shift of vacuum from $l$ to $l-1$ occurs around the point $x_l$.
Therefore $x_l$ becomes approximately the position of  a domain wall separating
 the vacua $l$ and $l+1$. \footnote{For a more detailed
 discussion of positions of walls, see Appendix A in \cite{INOS1}.}

As a slight modification of the above case, it is also possible to take the moduli
 matrix where there are nonzero components in the second line in (\ref{up})
 with zero in the  corresponding column components in the first line.
For example,
\ba
H_{0\la 1\leftarrow n \ra}=\left(
  \begin{array}{ccccccccc}
  \cdots & 0 & e^{r_1} & 0       & e^{r_2} & \cdots & e^{r_{n-1}} & 0 & \cdots \\
  \cdots & 0 & 0       & e^{r_n} & 0       & \cdots & 0           & 0 & \cdots
   \end{array}
 \right).\label{up-down}
\ea
This moduli matrix satisfies (\ref{const-Q4}).
However, a matrix where
 all elements are nonzero values only for one column, for example,
\ba
H_{0\la 1\leftarrow n \ra}=\left(
  \begin{array}{cccccccc}
  \cdots & 0 & e^{r_1} & e^{r_2} & \cdots & e^{r_{n-1}} & 0 & \cdots \\
  \cdots & 0 & e^{r_n} & 0       & \cdots & 0           & 0 & \cdots
   \end{array}
 \right), \label{up-down2}
\ea
is not allowed since it does not satisfy (\ref{const-Q4}). If there
is more than one column where all the components are nonzero values,
 the situation changes.
In other words, $H_0^{1 a}\neq 0$ and $H_0^{2 a}\neq 0$ with $a=1,\cdots n~(n\ge 2).$
As an example, we consider the following moduli matrix
\ba
H_{0\la 1\leftarrow 4 \ra}=\left(
  \begin{array}{cccccc}
  \cdots & 0 & e^{r_1} & e^{r_2} & 0 & \cdots \\
  \cdots & 0 & e^{r_4} & e^{r_3} & 0 & \cdots
   \end{array}
 \right).\label{multi-H}
\ea
The constraint (\ref{const-Q4}) gives
\ba
 e^{r_1+r_4}+e^{r_2+r_3}=0. \label{c-Q2}
\ea
There are four exponential factors and therefore it is expected to realize
 four vacua from this configuration:
\ba
&& H_{0\la 1 \ra}=
 \left(
  \begin{array}{cccccc}
   \cdots & 0 & 1 & 0 & 0 & \cdots \\
   \cdots & 0 & 0 & 0 & 0 & \cdots
  \end{array}
 \right),\quad
  H_{0\la 2 \ra}=
 \left(
  \begin{array}{cccccc}
   \cdots & 0 & 0 & 1 & 0 & \cdots \\
   \cdots & 0 & 0 & 0 & 0 & \cdots
  \end{array}
 \right), \no
&& H_{0\la 3 \ra}=
 \left(
  \begin{array}{cccccc}
   \cdots & 0 & 0 & 0 & 0 & \cdots \\
   \cdots & 0 & 0 & 1 & 0 & \cdots
  \end{array}
 \right),\quad
  H_{0\la 4 \ra}=
 \left(
  \begin{array}{cccccc}
   \cdots & 0 & 0 & 0 & 0 & \cdots \\
   \cdots & 0 & 1 & 0 & 0 & \cdots
  \end{array}
 \right). \label{vac-M}
\ea
However, because of (\ref{c-Q2}), only three vacua among them are realized
 from (\ref{multi-H}).
 In what follows, we will show this explicitly.

First of all, let us solve the constraint (\ref{c-Q2}).
For convenience, we introduce the notation $m_3\equiv -m_2$ and $m_4\equiv-m_1$,
 which is consistent with the inequivalent relation $m_A>m_{A+1}$ assumed before.
With the use of the relation (\ref{position}), (\ref{c-Q2}) is rewritten as
\ba
 e^{(m_1-m_2)X_1}+e^{(m_3-m_4)X_3}=0.
\ea
It is easily solved by
\ba
 X_1 = X_3 + {i(2n+1)\pi \over m_1-m_2},\quad n\in {\bf Z}. \label{sol-c}
\ea
It tells us
\ba
 x_1=x_3. \label{p-s}
\ea
Keeping the relation (\ref{p-s}) in mind, let us read vacua from (\ref{multi-H}).
Making the translation (\ref{shift}), we have
\ba
 H_{0\la 1\leftarrow 4 \ra}\rightarrow
 \left(
  \begin{array}{cccccc}
   \cdots & 0 & e^{m_1 x_0 +r_1} & e^{m_2 x_0 +r_2} & 0 & \cdots \\
   \cdots & 0 & e^{m_4 x_0 +r_4} & e^{m_3 x_0 +r_3} & 0 & \cdots
  \end{array}
 \right). \label{c-vac-H}
\ea
Further acting the world-volume symmetry transformation
 $V=e^{-m_2 x_0-r_2}$ on (\ref{c-vac-H}), we have
\ba
 H_{0\la 1\leftarrow 4 \ra}\rightarrow
 \left(
  \begin{array}{cccccc}
   \cdots & 0 & e^{(m_1-m_2)(x_0-X_1)} & 1 & 0 & \cdots \\
   \cdots & 0 & e^{-(m_2-m_3)(x_0-X_2)-(m_3-m_4)(x_0-X_3)} & e^{-(m_2-m_3)(x_0 - X_2)} & 0 & \cdots
  \end{array}
 \right). \label{c-vac-H2}
\ea
The upper-left and lower-right components are negligible if we consider the region
\ba
 x_1=x_3 \gg x_0 \gg x_2. \label{relation1}
\ea
By using (\ref{sol-c}), the lower-left component is written by
\ba -e^{-(m_2-m_3)(x_0-X_2)-(m_3-m_4)(x_0-X_1)}. \ea
Though the second term in the exponential is positive under the condition (\ref{relation1}),
 this exponential factor can be negligible if $x_2$ is taken to be small enough.
This condition can be still consistent with (\ref{relation1}). These
observations lead to the second vacuum labeled by $\la 2 \ra$
in (\ref{vac-M}).

Similarly, we can consider the third vacuum $\la 3 \ra$.
Multiplying (\ref{c-vac-H}) by $V=e^{-m_3 x_0-r_3}$, we have
\ba
H_{0\la 1\leftarrow 4 \ra}\rightarrow
 \left(
  \begin{array}{cccccc}
   \cdots & 0 & e^{(m_1-m_2)(x_0-X_1)+(m_2-m_3)(x_0-X_2)}  & e^{(m_2-m_3)(x_0-X_2)} & 0 & \cdots \\
   \cdots & 0 & e^{-(m_3-m_4)(x_0-X_3)} & 1& 0 & \cdots
  \end{array}
 \right). \label{c-vac-H3}
\ea
The upper-right and lower-left components can be negligible if we
consider the following region:
\ba
 x_2 \gg x_0 \gg x_3=x_1. \label{relation2}
\ea
It nevertheless conflicts with the condition
 (\ref{relation1}).
Therefore, the configuration (\ref{multi-H}) does not give
 the vacuum $\la 3 \ra$ in (\ref{vac-M}) in this case.
On the other hand, if (\ref{relation2}) holds,
 (\ref{multi-H}) gives the vacuum $\la 3 \ra$ while it does not give the vacuum $\la 2 \ra$.
We have now two possible parameter choices
\ba
&& x_1 = x_3 \gg x_2 \gg x_4, \label{c-1}\\
&& x_2 \gg x_1=x_3 \gg x_4, \label{c-2}
\ea
where the former (latter) leads to vacua in (\ref{vac-M}) except the
third (second) one. One can easily check that the first and fourth
vacua $\la 1\ra$ and $\la 4\ra$ in (\ref{vac-M}) are
 obtained from (\ref{multi-H}) with both the choices
(\ref{c-1}) and (\ref{c-2}).
Thus we find that (\ref{multi-H}) represents a double wall interpolating three vacua.

Finally we consider the odd $N$ case.
In this case, we have to take into account the scalar $H_0^{N+2}$.
The moduli matrices (\ref{up}) and (\ref{up-down}) are possible configurations
 with $H_0^{N+2}=0$, considering the constraint (\ref{const-Q4}),
 as in the single wall case.
The configuration (\ref{up-down2}) is also allowed
 if $H_0^{N+2}=\sqrt{2}i e^{(r_1+r_{n})/2}$.

The configuration (\ref{c-vac-H}) is also possible, but in this
case,
 (\ref{c-vac-H}) can give all the vacua (\ref{vac-M})
 with some parameter
 choices satisfying the constraint (\ref{c-Q2}) given by
\ba
 2(e^{r_1+r_4}+e^{r_2+r_3})+(H_0^{N+2})^2=0. \label{c-M2}
\ea
The solution (\ref{sol-c}) is not general any more.
 We can choose parameters by taking the nonzero value of $H_0^{N+2}$ so that $x_1 \gg x_2 \gg x_3 \gg x_4$,
 leading to all the vacua  (\ref{vac-M}).
Thus it is found that (\ref{multi-H}) with nonzero $H_0^{N+2}$
satisfying (\ref{c-M2})
 can represent a triple wall interpolating four vacua.

In this section, we have listed various possible forms of moduli
matrices. In what follows, by using the results here, we construct
an explicit solution for
 $N\le 4$ cases and investigate properties of walls.

%
%
\section{Explicit construction}
%
%
\setcounter{equation}{0}
\subsection{$N=1$ case}
In this case, there exist two vacua. The moduli matrix becomes
 2-component vectors $H_0^\a$ (index $a$ does not run)
 and one scalar $H_0^3$.
Here we form them as a
 3-component vector, $H_0^i=(H_0^1,H_0^2,H_0^3)$.
Moduli matrices exhibiting two vacua are given by
\ba
 & H_{0\langle 1 \rangle} = (1,0,0), \quad \sigma = -m, &\label{Q1-vac-H1} \\
 & H_{0\langle 2 \rangle}=(0,1,0), \quad \sigma = m,  & \label{Q1-vac-H2}
\ea
where $m$ is a mass parameter.
They satisfy the constraint (\ref{const-Q4}).
There should be only one domain wall connecting these two vacua.
Let us consider such a configuration.
We take a vector with a complex parameter $r$ as
\ba
 H_{0\langle 1 \leftarrow 2\rangle} = (1, e^{r}, H_0^3),\quad -\infty <{\rm Re}(r)<\infty,
 \label{h1}
\ea
where we choose the first component as a unit by using the
 world-volume symmetry transformation (\ref{equiv}). This configuration describes an
 elementary wall. The scalar $H_0^3$ is determined by the constraint
(\ref{const-Q4}) as
\ba
 H_0^3 = \sqrt{2}i e^{r/2}. \label{h3}
\ea
Repeating the same discussion as in the previous section of a single wall configuration
 for the odd $N$ case, it is found that the moduli matrix (\ref{h1}) gives
 the first vacuum (\ref{Q1-vac-H1}) at $x=\infty$ and the second
 vacuum (\ref{Q1-vac-H2}) at $x=-\infty$.

Once the moduli matrix $H_0$ is specified, the BPS wall solution is easily derived.
Substituting (\ref{h1}) into (\ref{const-Q3}), we have
\ba
 S=\sqrt{e^{2m x}+e^{-2m x + 2{\rm Re}(r)}+2e^{{\rm Re}(r)}}. \label{h2}
\ea
Therefore, from (\ref{b4}), we obtain the solution
\ba
 && \displaystyle
   \F^{1} = {e^{m(x-x_1)} \over \sqrt{e^{2m(x-x_1)}+e^{-2m (x - x_1)}+2}},\\
 && \displaystyle
   \F^{2} = {e^{-m(x-x_1)+i{\rm Im}(r)} \over \sqrt{e^{2m(x-x_1)}+e^{-2m (x - x_1)}+2}},  \\
 && \displaystyle
   \phi^{3} = {\sqrt{2}ie^{i{\rm Im}(r)/2} \over \sqrt{e^{2m (x-x_1)}+e^{-2m (x - x_1)}+2}},
\ea
where $x_1={\rm Re}(X_1)=-r/m$ as defined in (\ref{position}). The imaginary part of  $r$ is a moduli with respect to
  the broken $SO(2)$ phase. We can also obtain $\sigma$ through (\ref{b1}).
Gauging away $v_1$ from (\ref{b1}) by using the $U(1)$ gauge transformation and
 substituting (\ref{h2}) into (\ref{b1}), we have
\ba
 \sigma = -{m(e^{2m(x-x_1)} - e^{-2m(x-x_1)}) \over e^{2m(x-x_1)} + e^{-2m(x-x_1)}+2}.
\ea

\begin{figure}[h!]
\begin{center}
 $
  \begin{array}{ccc}
  \epsfxsize=7cm
   \epsfbox{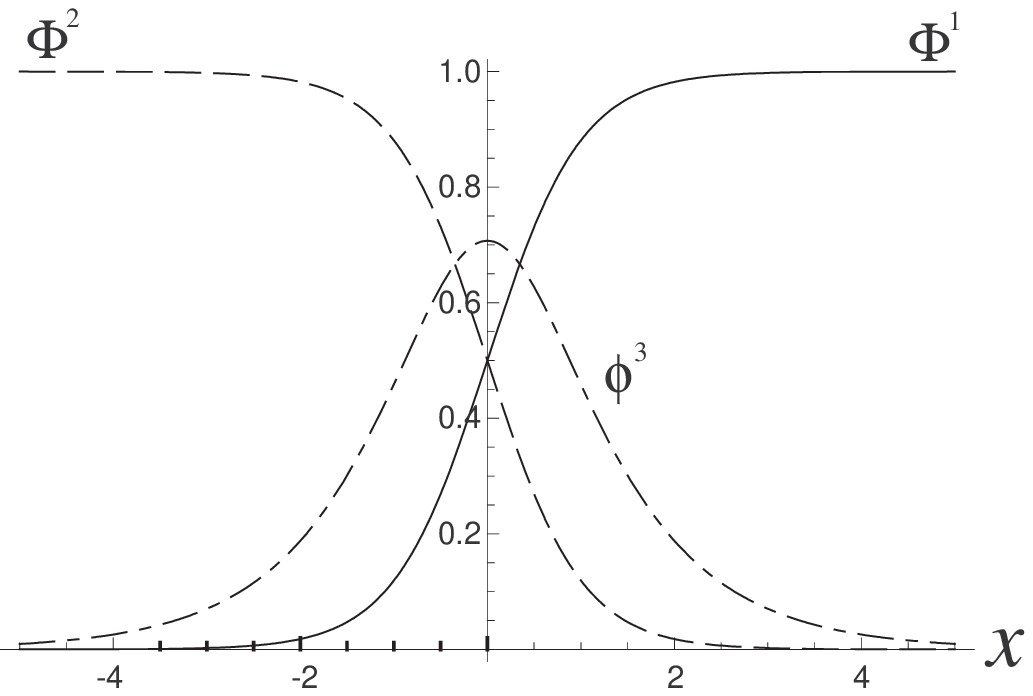} &
                      &
  \epsfxsize=7cm
   \epsfbox{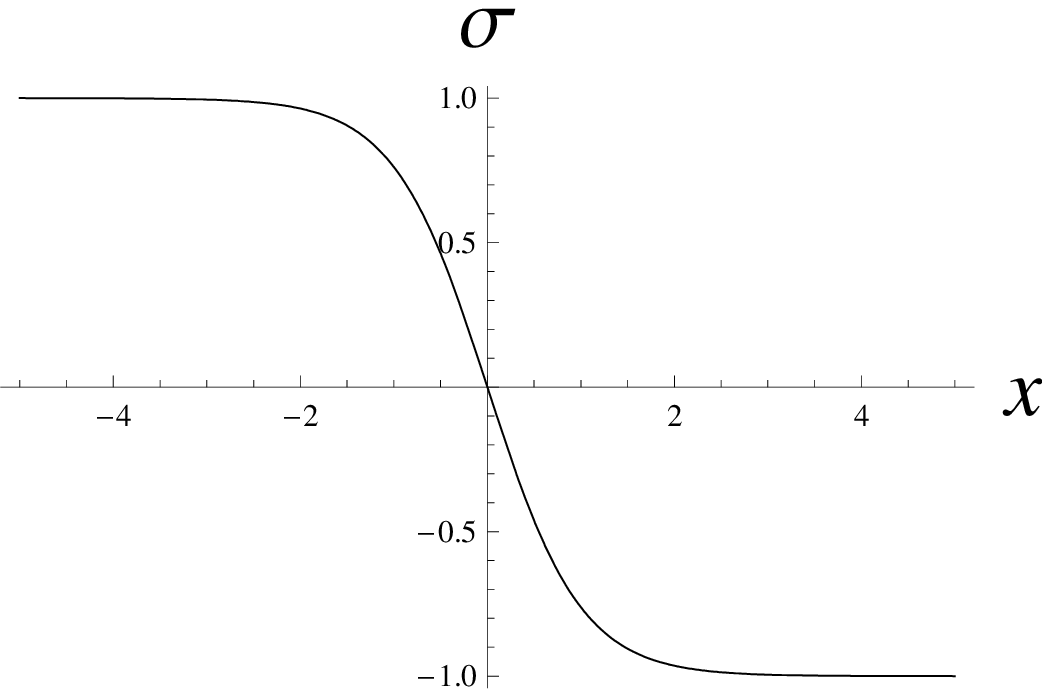}
  \end{array}
 $
\caption{Plots of $\F^1$(solid curve in the left figure),
 $\F^2$(dashed curve), {\rm Re}$(\phi^3)$(dot-dashed curve) and
 $\sigma$(solid curve in the right figure) with $m=1$ and $r=0$.}
\label{Q1}
\end{center}
\end{figure}

Plots of these configurations are shown in Fig. \ref{Q1}.
The configurations of $\F^1$, $\F^2$ and $\sigma$ form one domain wall solution
 while $\phi^3$ is a solution connecting two trivial vacua.
They give the first vacuum (\ref{Q1-vac-H1}) in the limit
 $x\rightarrow \infty$ and the second vacuum (\ref{Q1-vac-H2}) in the limit
 $x\rightarrow -\infty$.
From the plots, it is also seen that $x_1$ defined by (\ref{position}) is actually
 the position of the wall.

As we mentioned in the end of Section \ref{sec;massiveKahler}, the
number of vacua in the $Q^1$ case is
 consistent with the result in the massive NLSM on $T^*{\bf C}P^1$.
The latter gives one domain wall solution. Therefore, our wall
 solution is also consistent with this. However, in the $Q^1$ case,
 there is also a solution expressed by $\phi^3$
 in addition to the domain wall solution.
This does not exist in the massive NLSM on $T^*{\bf C}P^1$\cite{ANNS}.
The difference
 stems from a different parametrization of the two models.
The nonzero solution $\phi^3$ is necessary to obtain the wall
 solution in the $Q^1$ case. Vanishing $\phi^3$ means that $H_0^3=0$
 in (\ref{h1}). The moduli matrix (\ref{h1}) with $H_0^3=0$ does not
 satisfy the constraint (\ref{const-Q4})
 and it is therefore no longer a solution.

%
%
\subsection{$N=2$ cases} \label{sec;N=2}
In this case, there exist four vacua.
Therefore,
 there should exist richer configurations such as multiwall solutions.
The configuration is described by the moduli matrix
 written by a 2 times 2 matrix  $H_0^{\a a}$($\a=1,2$, $a=1,2$).
Moduli matrices corresponding to four vacua are given by
\ba
H_{0\la 1 \ra}=
  \left(
  \begin{array}{cc}
   1 & 0 \\
   0 & 0
  \end{array}
 \right),~~
H_{0\la 2 \ra}=
 \left(
  \begin{array}{cc}
   0 & 1  \\
   0 & 0
  \end{array}
 \right),~~
H_{0\la 3 \ra}=
 \left(
  \begin{array}{cc}
   0 & 0 \\
   0 & 1
  \end{array}
 \right),~~
H_{0\la 4 \ra}=
 \left(
  \begin{array}{cc}
   0 & 0 \\
   1 & 0
  \end{array}
 \right). \label{Q2-vac-H}
\ea

Single wall configurations are easily obtained,
 following the classification of single walls in Section \ref{sec;moduli matrix}.
It is found that there are four possible single wall configurations given by
\ba
&& H_{0\la 1 \leftarrow 2 \ra}=
 \left(
  \begin{array}{cc}
   1 & e^r \\
   0 & 0
  \end{array}
 \right),~~
  H_{0\la 1 \leftarrow 3 \ra}=
 \left(
  \begin{array}{cc}
   1 & 0 \\
   0 & e^r
  \end{array}
 \right),~~\label{s-a}\\
&&  H_{0\la 2 \leftarrow 4 \ra}=
 \left(
  \begin{array}{cc}
   0 & 1 \\
   e^r & 0
  \end{array}
 \right),~~
 H_{0\la 3 \leftarrow 4 \ra}=
 \left(
  \begin{array}{cc}
   0 & 0 \\
   e^r & 1
  \end{array}
 \right), \label{single-Q2}
\ea
where $-\infty<{\rm Re}(r)<\infty$. We recognize  that they are all
elementary walls. These lead to corresponding vacua in
(\ref{Q2-vac-H}) at boundaries, $x=\pm \infty$. Here we have taken
one of the exponential factors to be a unit by using (\ref{equiv}).
Moduli matrices
\ba
 H_{0\la 1 \leftarrow 4 \ra}=
 \left(
  \begin{array}{cc}
   1 & 0 \\
   e^r & 0
  \end{array}
 \right),~~
  H_{0\la 2 \leftarrow 3 \ra}=
 \left(
  \begin{array}{cc}
   0 & 1 \\
   0 & e^r
  \end{array}
 \right), \label{Q2-not}
\ea
 are not allowed since they do not satisfy the constraint (\ref{const-Q4}).
Explicit solutions for $\F^{\a a}$ and $\sigma$ are obtained
 from (\ref{s-a}) and (\ref{single-Q2}), as obtained in the $Q^1$ case.

Next we consider a multiwall solution.
A possible form of moduli matrix for such a configuration is given by
\ba
 H_{0\la 1\leftarrow 4\ra}=
 \left(
  \begin{array}{cc}
   e^{r_1} & e^{r_2} \\
   e^{r_4} & e^{r_3}
  \end{array}
 \right), \label{Q2-m}
\ea
where we take $r_1=0$ by using (\ref{equiv}).
This is exactly the same case as in (\ref{multi-H}).
As we have seen in the discussion below (\ref{multi-H}),
 (\ref{Q2-m}) realizes a double wall configuration interpolating three vacua.
This is the multiwall composed of the maximal number of single walls in $Q^2$ case.

Using the same definition of (\ref{position}), we have two possible parameter
 choices (\ref{c-1}) and (\ref{c-2}).
The former choice gives a configuration interpolating the vacua
 labeled by $\la 1 \ra$, $\la 2 \ra$ and $\la 4 \ra$
 and the latter gives one interpolating the vacua
 labeled by $\la 1 \ra$, $\la 3 \ra$ and $\la 4 \ra$.
In order to make clear which vacua are interpolated,
 in what follows, we shall write these configurations
 as $H_{0\la 1\leftarrow 2 \leftarrow 4\ra}$
 and $H_{0\la 1\leftarrow 3 \leftarrow 4\ra}$, respectively.
An explicit solution is given by
\ba
 \Phi^{\a a}=
  {1 \over \left(\displaystyle \sum_{i=1}^4e^{2m_i x + 2{\rm Re}(r_i)}\right)^{1/2}}
  \left(
  \begin{array}{cc}
   e^{m_1 x + r_1} & e^{m_2 x + r_2} \\
   e^{m_4 x + r_4} & e^{m_3 x + r_3}
  \end{array}
 \right), \quad
 \sigma = -{\displaystyle \sum_{i=1}^4 2 m_i e^{2 m_i x + 2{\rm Re}(r_i)}
  \over \displaystyle \sum_{i=1}^4e^{2m_i x + 2{\rm Re}(r_i)}}, \label{dw}
\ea
where $m_3\equiv -m_2$ and $m_4\equiv -m_1$, and the complex parameter
 $r_i~(i=1,\cdots,4)$ should satisfy the constraint (\ref{c-Q2}).

From the double wall configuration one can obtain a single wall configuration (\ref{s-a}) and
 (\ref{single-Q2}).
For instance, taking the limit of $r_3\rightarrow -\infty$ and $r_4\rightarrow -\infty$ in (\ref{Q2-m}),
 and using the world-volume symmetry transformation (\ref{equiv}),
 it reduces to the single wall configuration $H_{0\la 1\leftarrow 2\ra}$.
The constraint (\ref{const-Q4}) becomes trivial in this limit. Note
that by taking this limit, boundary conditions at $x=\pm \infty$ are
 changed from $\la 1\leftarrow 4 \ra$ to $\la 1 \leftarrow 2\ra$. The
 physical meaning of this transition is that one of walls labeled by
 $\la 2\leftarrow 4 \ra$ in the double wall
 moves away to infinity
 along the $x$ direction and the other one labeled by $\la 1 \leftarrow 2\ra$ is left.
Similarly, one can recover all the configurations in (\ref{s-a}) and (\ref{single-Q2}) from (\ref{Q2-m}).

\begin{figure}[h!]
\begin{center}
 $\begin{array}{ccc}
  \epsfxsize=5cm
   \epsfbox{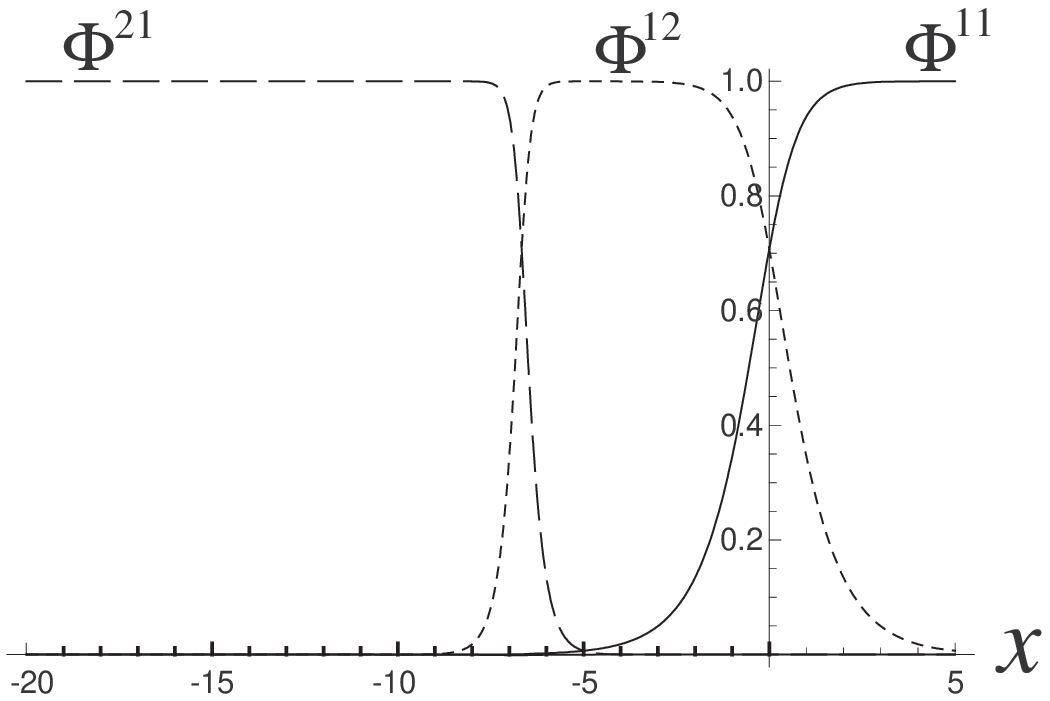}
  &
   \epsfxsize=5cm
   \epsfbox{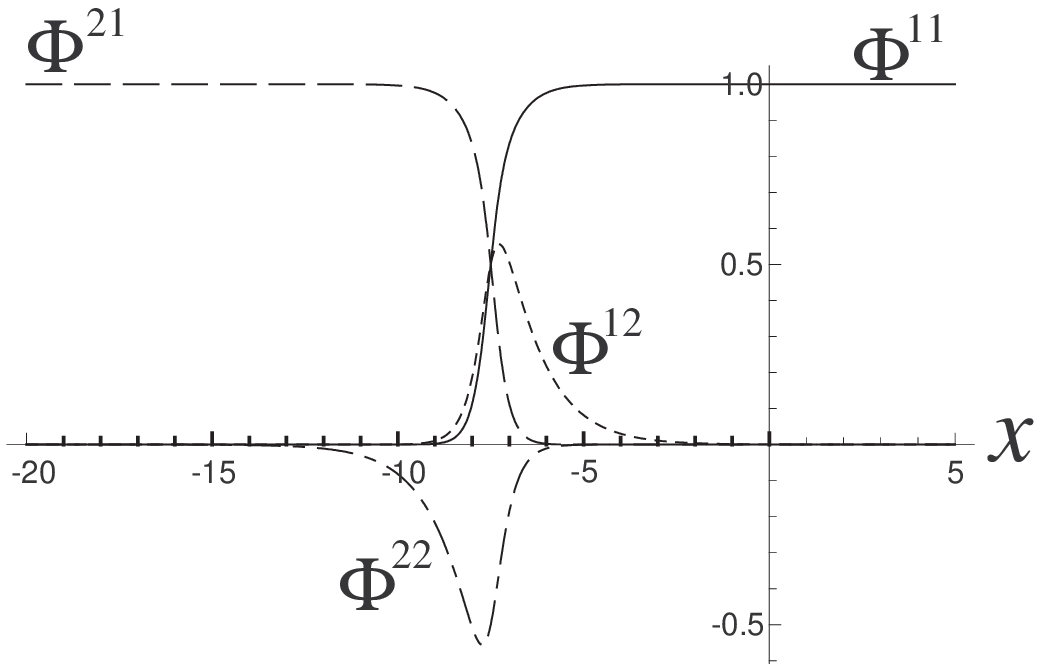}
  &
   \epsfxsize=5cm
   \epsfbox{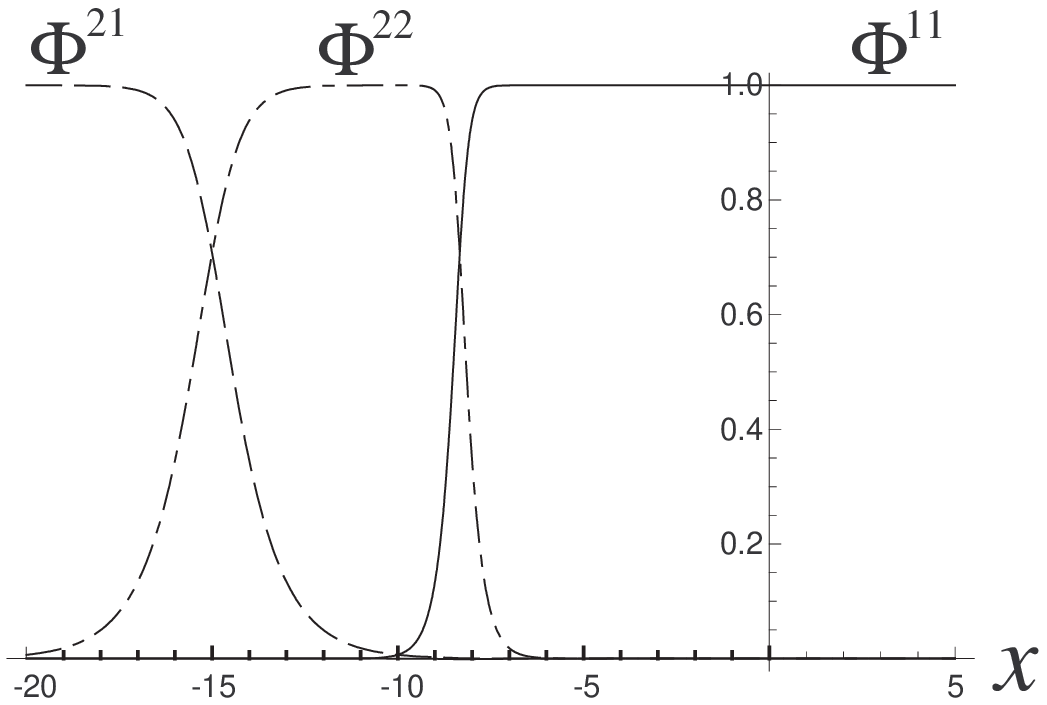}
  \\
   \epsfxsize=5cm
   \epsfbox{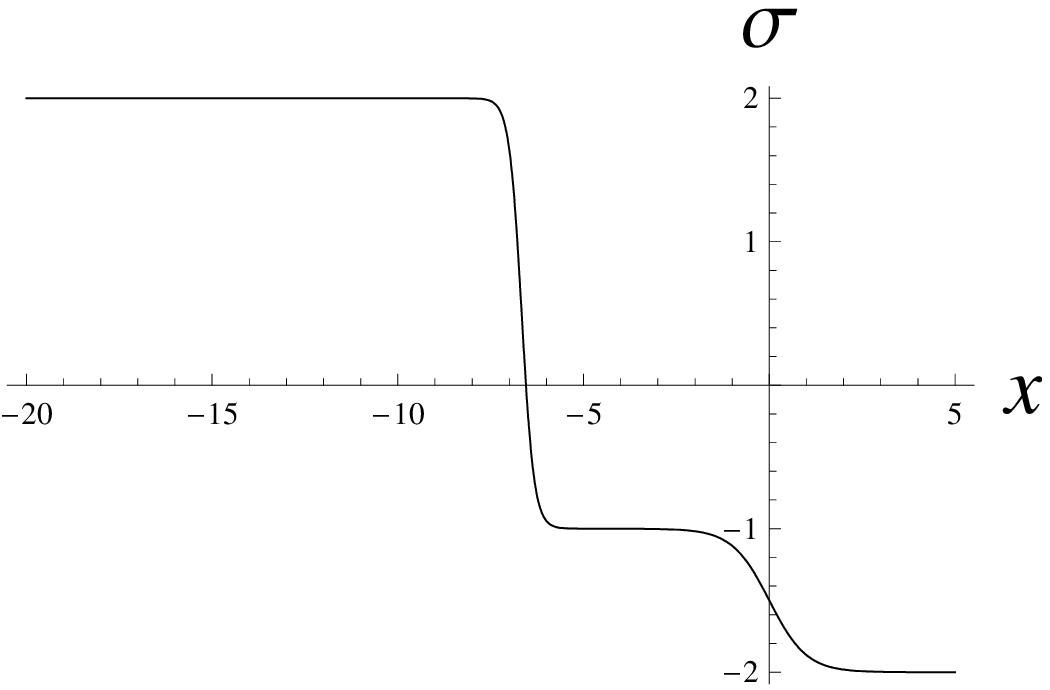}
  &
   \epsfxsize=5cm
   \epsfbox{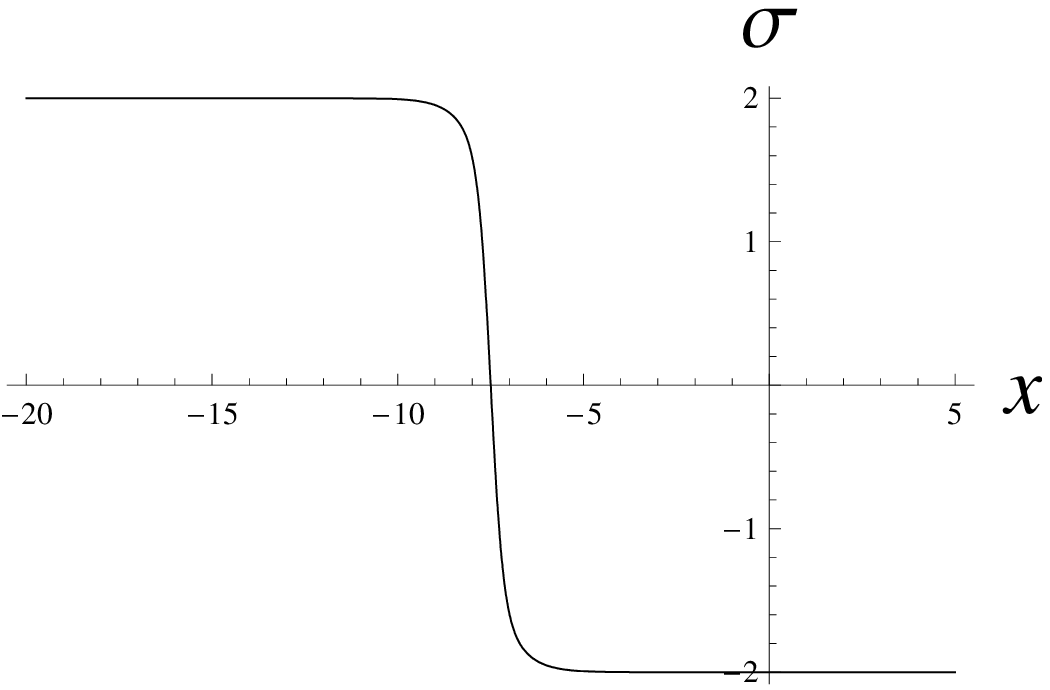}
  &
   \epsfxsize=5cm
   \epsfbox{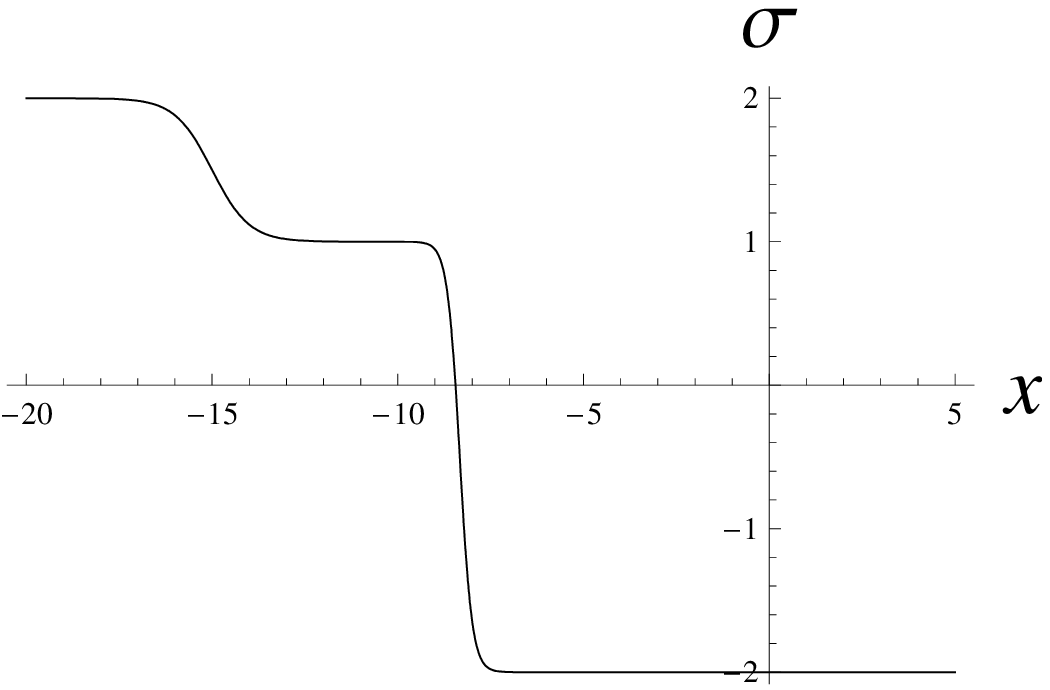}
    \\
   \epsfxsize=5cm
   \epsfbox{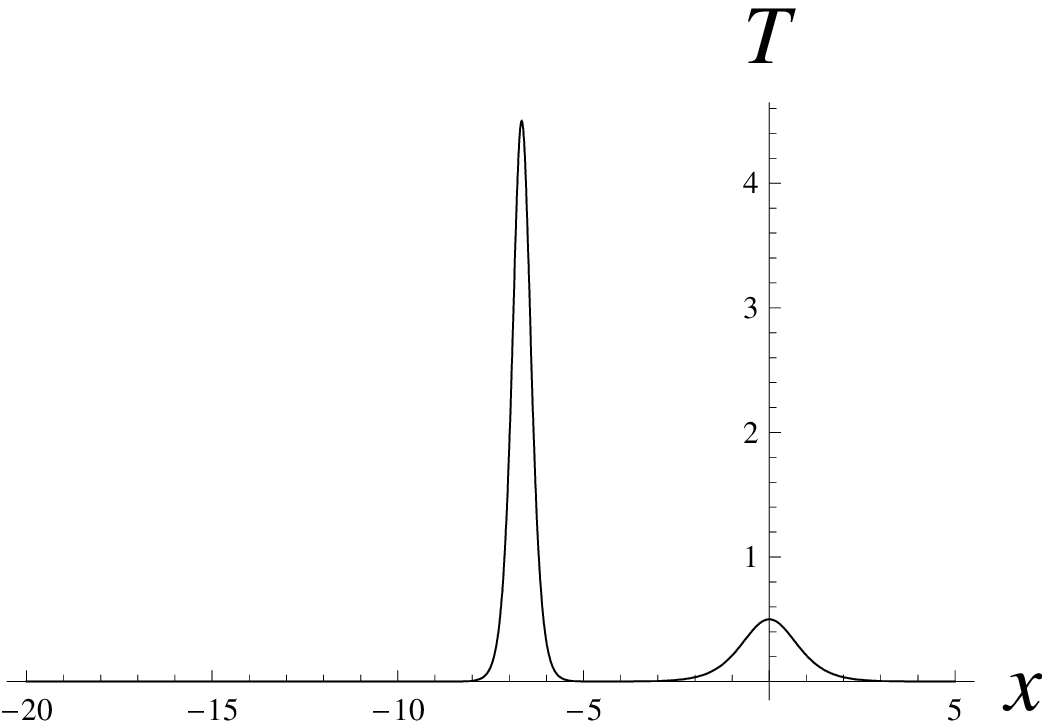}
  &
   \epsfxsize=5cm
   \epsfbox{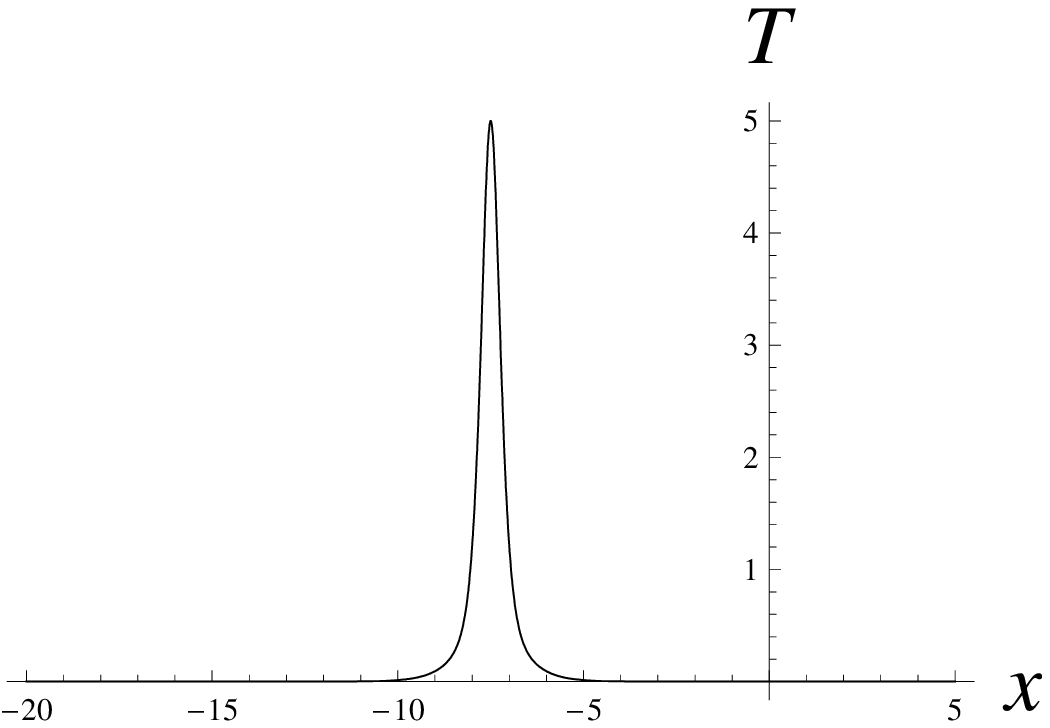}
  &
   \epsfxsize=5cm
   \epsfbox{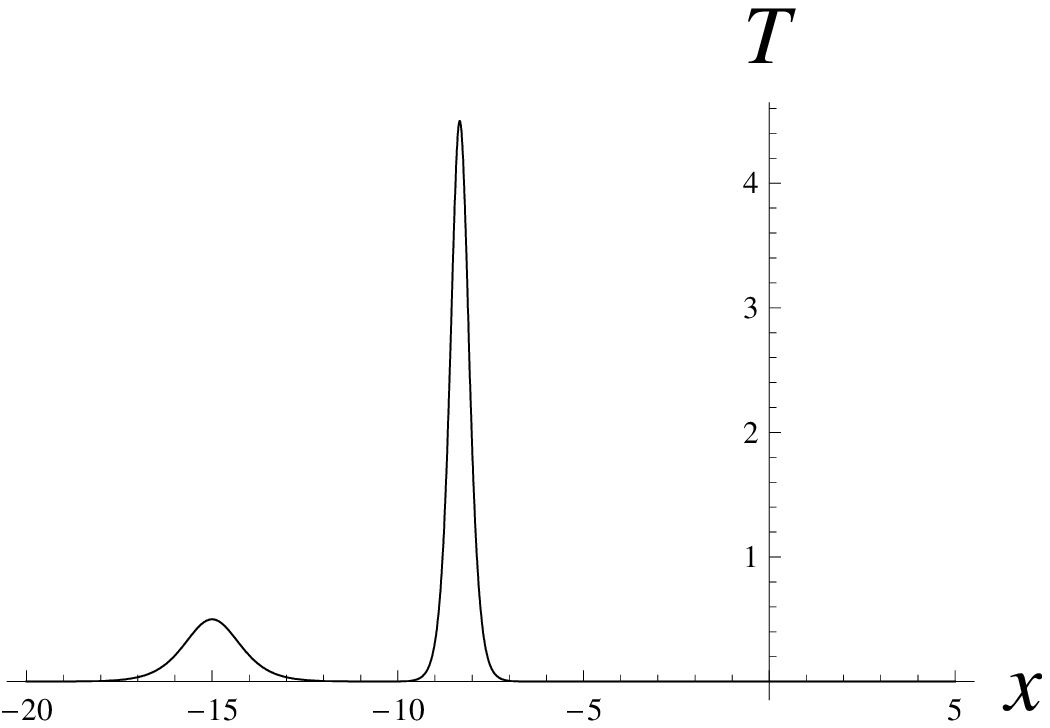}
  \end{array}
$ \caption{Plots for double wall configurations. The left figures
from the top to bottom show plots of $\Phi$, $\sigma$ and $T$
 for the region (\ref{c-1}) with
 $r_1=0,~r_2=0,~r_3=-20,~r_4=-20+i\pi$, respectively.
The middle and the right figures show the same plots
 for the parameter region $x_1=x_2=x_3>x_4$ with $r_1=0,~r_2=-15/2,~r_3=-45/2,~r_4=-30$
 and (\ref{c-2}) with $r_1=0,~r_2=-15,~r_3=-25+i\pi,~r_4=-40$, respectively.
Solid, dotted, dashed and dot-dashed curves in the top figures
depict $\F^{11},~\F^{12}$,
 $\F^{21}$ and $\F^{22}$, respectively.
For all plots we take $m_1=2$ and $m_2=1$.} \label{Q2-multi}
\end{center}
\end{figure}

\begin{figure}[h!]
\begin{center}
$\begin{array}{ccc}
  \epsfxsize=6.5cm
   \epsfbox{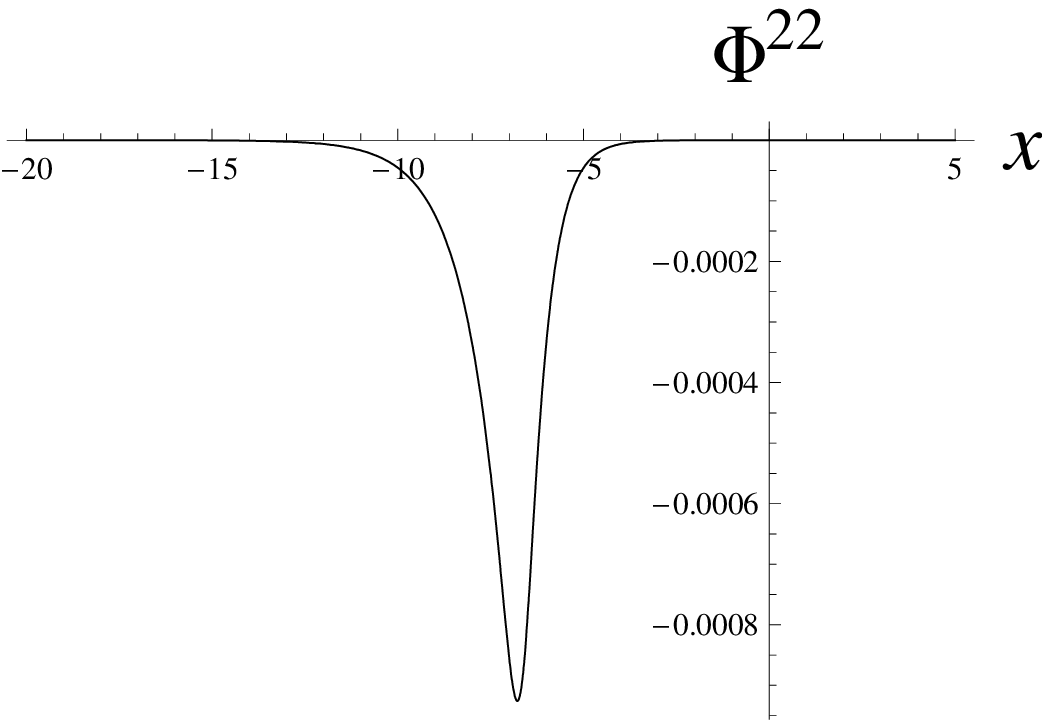}
  & &
    \epsfxsize=6.5cm
   \epsfbox{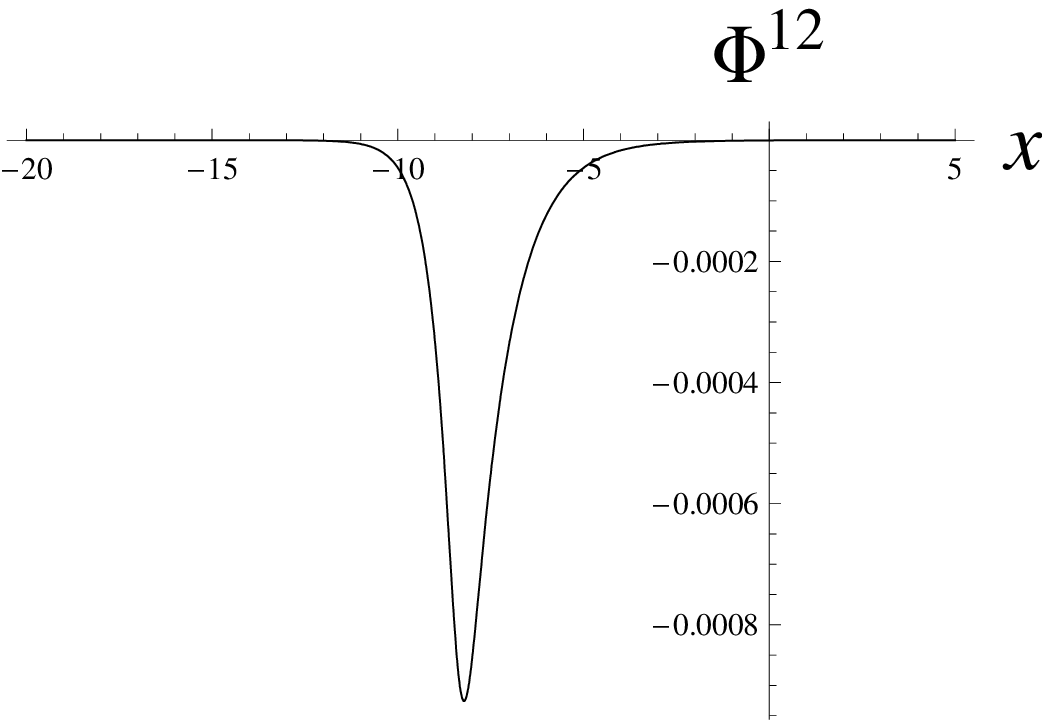}
  \end{array}
$ \caption{Plots for $\Phi^{22}$ and $\F^{12}$ for the regions
(\ref{c-1})
 and (\ref{c-2}) with the same parameter choices in the left and the right figures
 in Fig. \ref{Q2-multi}.}
\label{Q2-lump}
\end{center}
\end{figure}

The double wall configuration (\ref{dw}) has another remarkable
property. By varying  the moduli parameters $r_i$, the configuration
$H_{0\la 1\leftarrow 2 \leftarrow 4\ra}$
 can be obtained from $H_{0\la 1\leftarrow 3 \leftarrow 4\ra}$ and vice versa.
Through this transition, a pair of walls in the configuration
 commutes each other. In Fig. \ref{Q2-multi}, we illustrate this
 phenomenon. The left figures in Fig. \ref{Q2-multi} depict the plots
 of $\F,~\sigma$
 and the tension $T$ for the parameter region (\ref{c-1}) from top to bottom,
 corresponding to the configuration $H_{0\la 1 \leftarrow 2 \leftarrow 4\ra}$.
Three vacua $\la 1 \ra$, $\la 2 \ra$ and $\la 4 \ra$ are interpolated by
 $\F^{11},~\F^{12}$ and $\F^{21}$ with $\sigma$.
The component $\F^{22}$ just interpolates trivial vacua at
 both boundaries and whose absolute value is much less than 1 (see Fig.
 \ref{Q2-lump}). Two walls approach as $r_i$ varies appropriately and
 are located at the same position as
 $x_1=x_2=x_3>x_4$ (middle figures in Fig. \ref{Q2-multi}).
In this parameter region,
 not only $\F^{22}$ but also $\F^{12}$ have absolute values less than 1.
It means that the configuration does not interpolate the vacuum $\la 2 \ra$ anymore.
The components $\F^{11}$ and $\F^{21}$ with $\sigma$ only interpolate the vacua $\la 1 \ra$
 and $\la 4 \ra$ and form a single wall.
Taking parameters which satisfy (\ref{c-2}), the absolute value of
 $\F^{22}$ gets increased
 up to 1 while one of $\Phi^{12}$ decreases further (see
 Fig. \ref{Q2-lump}). As a result, the configuration can interpolate the
 vacua labeled by $\la 1 \ra$, $\la 3 \ra$
 and $\la 4 \ra$ (right figures in Fig. \ref{Q2-multi}).
Clearly it corresponds to the configuration $H_{0\la 1 \leftarrow 3 \leftarrow 4\ra}$.
Here the signs of $\F^{12}$ and $\F^{22}$ flip from the previous two cases,
 which stem from the solution of the constraint (\ref{sol-c}).
Now we have seen that as a pair of walls commutes each other,
 the intermediate vacua $\la 2 \ra$ and $\la 3 \ra$ exchange, keeping the vacua at boundaries
 $\la 1 \ra$ and $\la 4\ra$ unchanged.
It has been first observed in a SUSY $U(N_C)$ gauge theory which
 is coupled to $N_F$ massive flavors in the presence of
 the Fayet-Iliopoulos term with eight supercharges
 \cite{INOS1}.
They are called the penetrable walls.
 In \cite{INOS1}, this phenomenon appears as a non-Abelian nature.
(Especially the case for
 $N_C=2$ and $N_F=4$ has been investigated there.)
However, here it is found that the penetrable walls are also
possible in Abelian case since
 our model is based on the Abelian gauge theory.

\begin{figure}[h!]
\begin{center}
 $\begin{array}{ccc}
  \epsfxsize=6cm
   \epsfbox{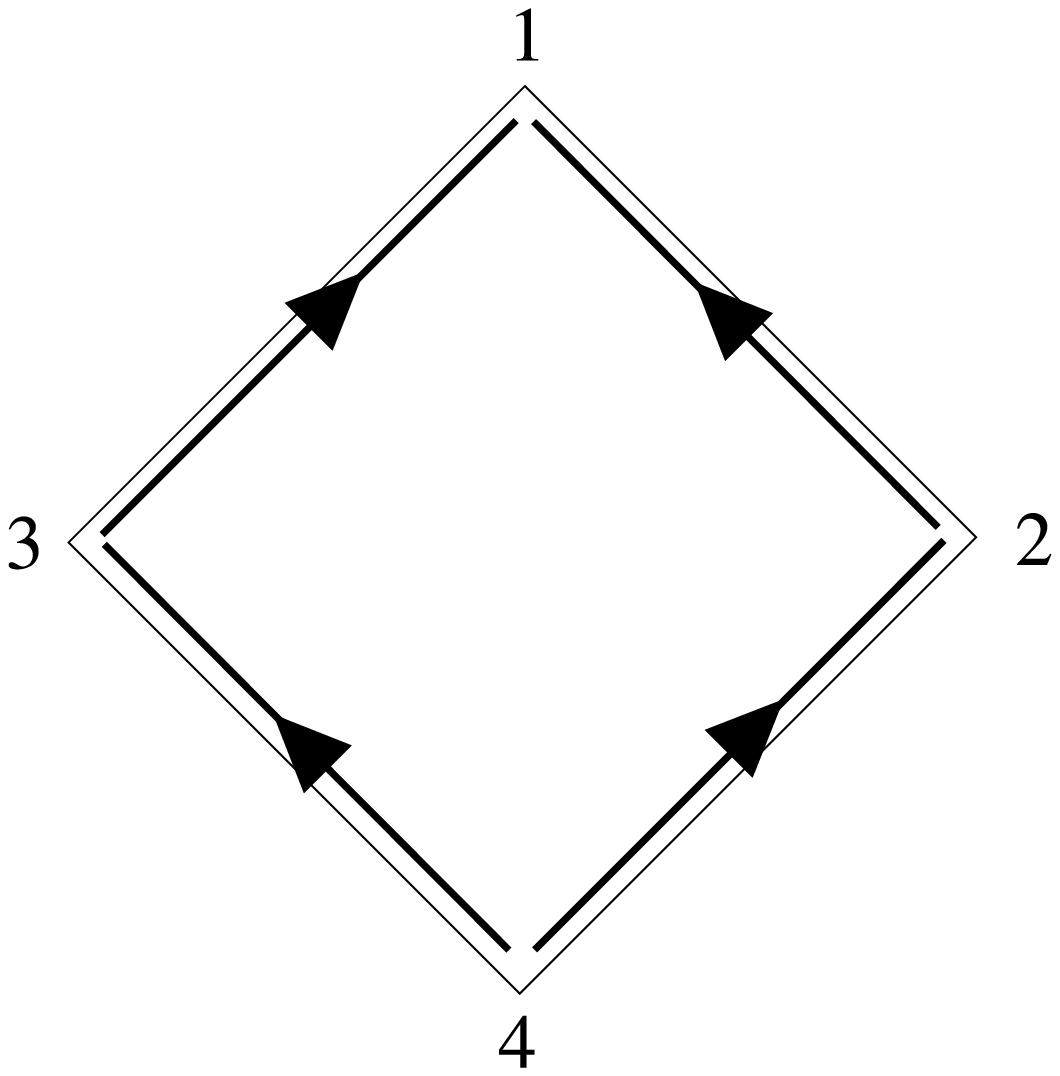}
  & &
  \epsfxsize=6cm
   \epsfbox{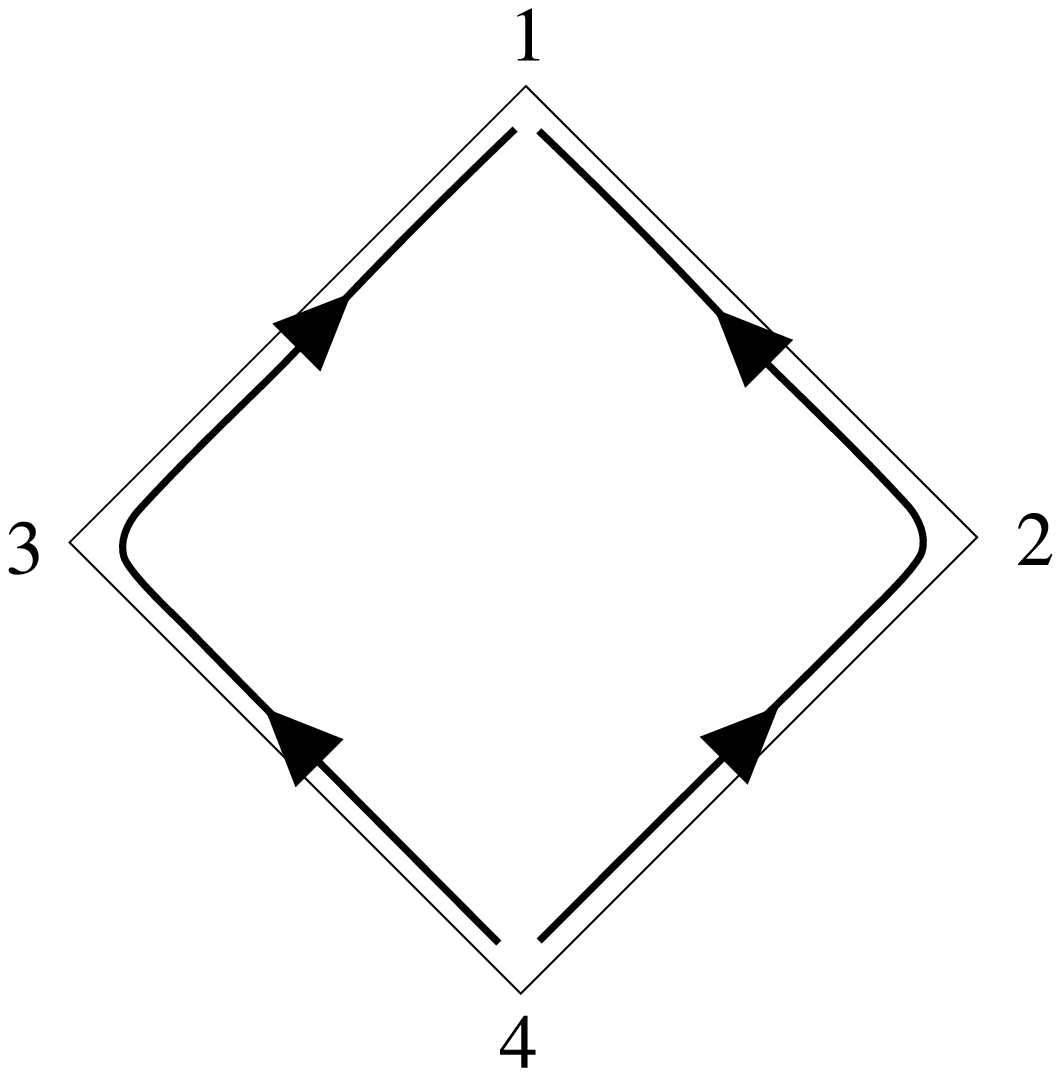}
  \end{array}
 $
\caption{All single (left) and double (right) wall configurations
for the $Q^2$ case.
 A number labels a vacuum.
 An arrow with an arrowhead denotes an elementary single wall in the left figure.
 A single line with two arrowheads denotes a double wall composed of two elementary
 walls in the right figure.}
\label{Q2-summary}
\end{center}
\end{figure}

In Fig. \ref{Q2-summary}, we show diagrams representing all possible configurations
 for $Q^2$ which consist of single and double wall solutions.
It is found that there are four elementary walls and two double walls.

%
%
\subsection{$N=3$ case}
In this case,
 the theory has the same number of vacua as the $N=2$ case.
The moduli matrix is also written as the $N=2$ case, $H_0^{\a
 a}(\a=1,2,~a=1,2)$, but there
 is an additional scalar $H_0^5$.
Even though there are the same number of vacua as the $N=2$ case,
possible
 wall configurations are more abundant than the $N=2$ case.
The main reason comes from the difference of the constraint
(\ref{const-Q4}), in which
 $H_0^5$ comes in.
Because of this, for instance, there can be a triple wall which is
 absent in the $N=2$ case. In the following, we will list possible
 configurations and discuss properties of solutions.

Moduli matrices representing
 vacua have the same form with (\ref{Q2-vac-H}).
The scalar $H_0^5$ is determined by (\ref{const-Q4}), yielding $H_0^5=0$.

\begin{figure}[h!]
\begin{center}
 $\begin{array}{ccc}
  \epsfxsize=5cm
   \epsfbox{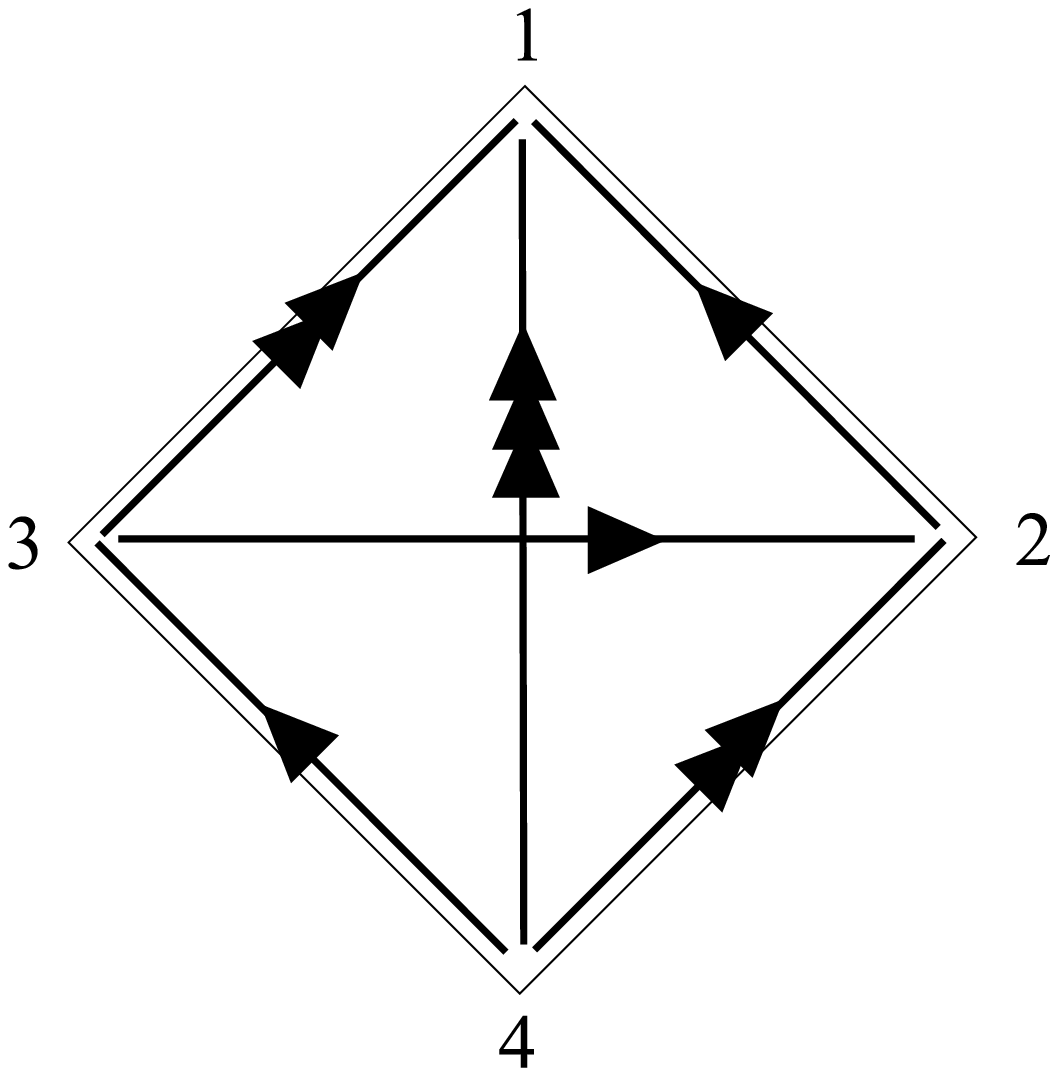}
  &
  \epsfxsize=5cm
   \epsfbox{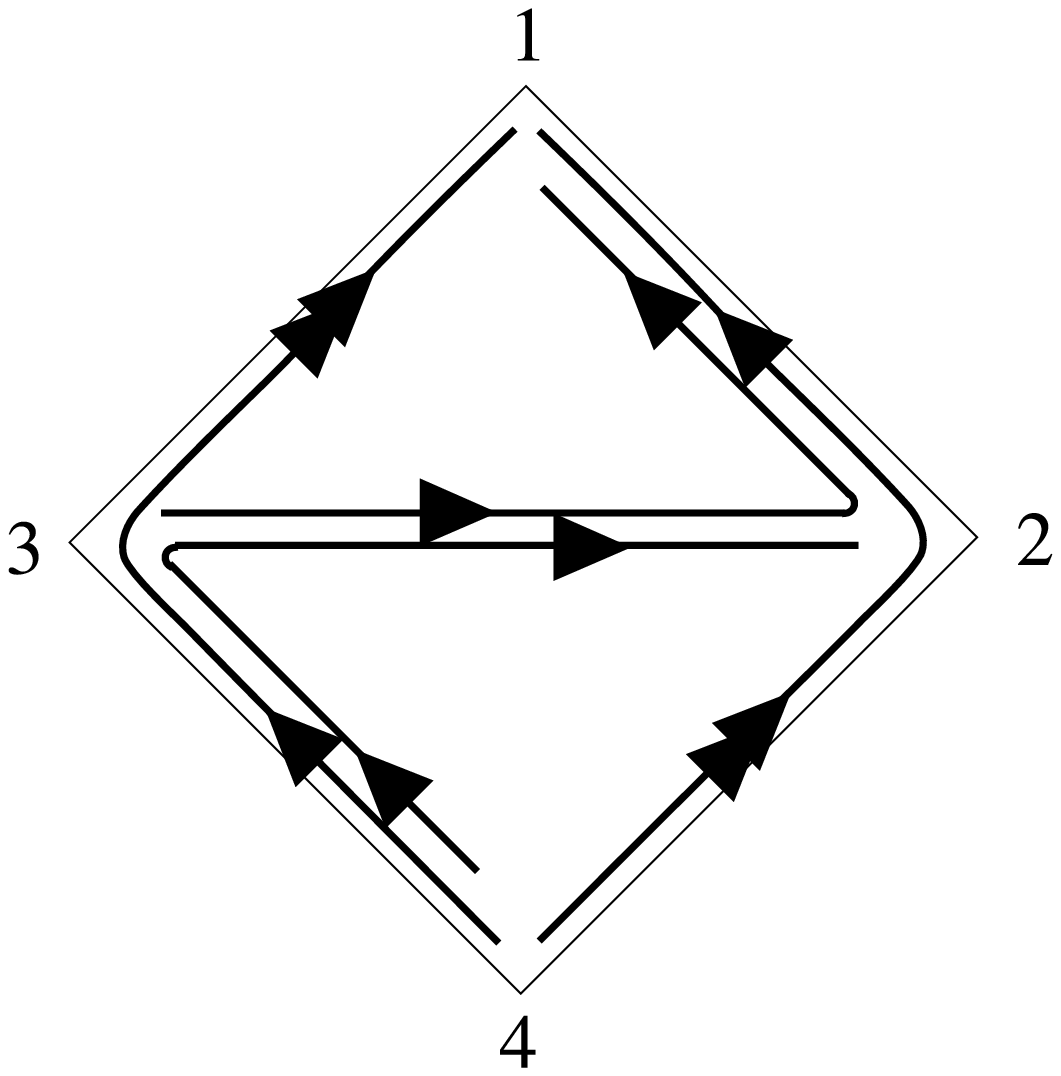}
  &
  \epsfxsize=5cm
   \epsfbox{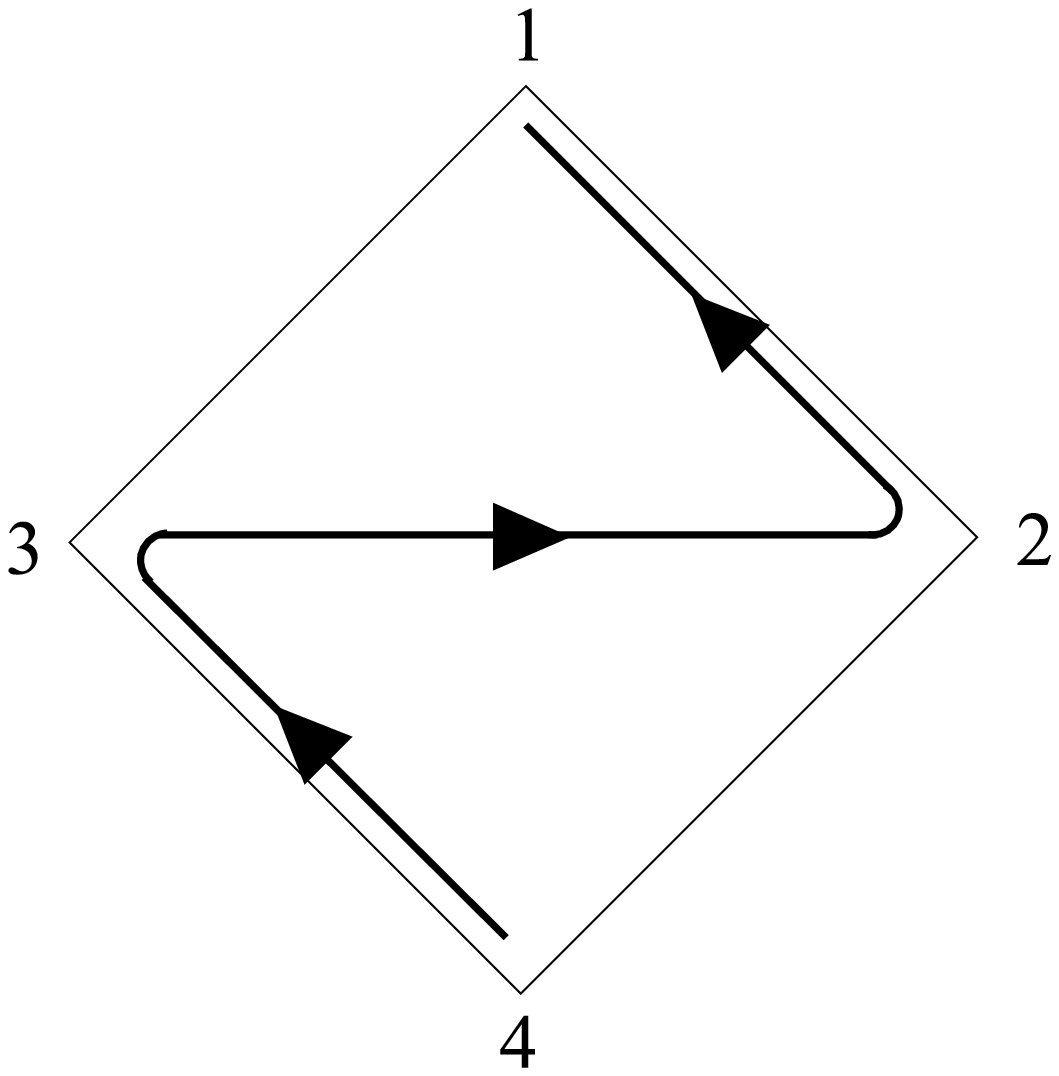}
  \end{array}
 $
\caption{All single (left), double (middle) and triple (right) wall
configurations
 for $Q^3$ case.
An arrow with an arrowhead denotes an elementary single wall and one
with
 a two(three) arrowheads denotes a compressed single wall of level one(two)
 in the left figure.
In the middle figure, there are two double walls composed of two
elementary walls
 and two double walls composed of one elementary wall and one compressed wall of
 level one.
A triple wall in the right figure is composed only by elementary
walls. } \label{Q3-summary}
\end{center}
\end{figure}

Single wall configurations existing in the $N=2$ case, (\ref{s-a})
and (\ref{single-Q2}),
 are possible if $H_0^5=0$.
In addition, as discussed below (\ref{odd-H}), there are two possible single
 wall configurations
\ba
&H_{0\la 1\leftarrow 4 \ra}=\pmatrix { 1 & 0\cr e^r& 0},\quad H_0^5=\sqrt{2} i e^{r/2}, & \label{k}\\
&H_{0\la 2\leftarrow 3 \ra}=\pmatrix {0&1 \cr 0 & e^r},\quad H_0^5=\sqrt{2}i e^{r/2}. & \label{l}
\ea
Explicit forms of solutions for $\F^{\a a}$ and $\sigma$ are
obtained as in the $Q^1$ case. Putting (\ref{s-a}),
(\ref{single-Q2}), (\ref{k}) and (\ref{l}) together,
 there are six single domain walls (see also the left figure in Fig. \ref{Q3-summary}).
%
Some of them are elementary walls, but the others are compressed
 walls. The latter does not occur in the $N=2$ case.

Let us discuss a double wall configuration. There are two types of
 moduli matrices to express a double wall configuration. One form is
 exactly the same as (\ref{Q2-m}) in the $N=2$ case where all
 components in $H_0^{\a a}$ are nonzero values, and the scalar
 $H_0^5$ is given by $H_0^5=0$. By this setting, all the discussions
 are the same as in the $N=2$ case.

Another form is
 given by $H_0^{\a a}$
 where there are three nonzero components and the nonzero $H_0^5$.
The latter is determined by (\ref{const-Q4}). Forms of moduli
 matrices in this case are listed below:
\ba
&&H_{0 \la 1\leftarrow 2 \leftarrow 3 \ra}
 =\pmatrix {1&e^{r_2}\cr 0& e^{r_3}}, \quad H_0^5=\sqrt{2}ie^{(r_2+r_3)/2}, \label{a} \\
&&H_{0 \la 2\leftarrow 3\leftarrow 4 \ra}=\pmatrix {0 &1 \cr e^{r_4} & e^{r_3}},
 \quad H_0^5=\sqrt{2}i e^{r_3/2}, \label{b} \\
&&H_{0 \la 1 \leftarrow 2\leftarrow 4 \ra}=\pmatrix {1& e^{r_2}\cr e^{r_4} &0},
 \quad H_0^5=\sqrt{2}ie^{r_4/2}, \label{c} \\
&&H_{0 \la 1\leftarrow 3 \leftarrow 4 \ra}=\pmatrix { 1& 0 \cr e^{r_4}&e^{r_3}},
 \quad H_0^5=\sqrt{2}i e^{r_4/2}. \label{d}
\ea
Here again we take one of the components to be a unit by using
(\ref{equiv}). From these configurations, we can obtain an
elementary wall and a compressed wall. For example, let us consider
the configuration (\ref{a}). If one takes the limit of $r_3
\rightarrow -\infty$,
 the configuration reduces to
\ba
H_{0 \la 1 \leftarrow 2 \ra}=\pmatrix{1 & e^{r_2} \cr 0 & 0},\quad H_0^5=0. \label{a1}
\ea
The boundary condition changes from $\la 1 \leftarrow 3 \ra$ to $\la 1 \leftarrow 2 \ra$.
It means that by this limit one of walls
 labeled by
 $\la 2 \leftarrow 3 \ra$
 goes away to infinity along the $x$ direction.
Therefore, this limit realizes an elementary wall
 labeled by $\la 1 \leftarrow 2\ra$.

Next we consider another limit.
First we multiply (\ref{a}) by $V=e^{-r_2}$ by using the world-volume
 symmetry transformation (\ref{equiv}).
Then taking the limit of $r_2\rightarrow \infty$, with keeping the parameter $r_3-r_2$
 finite, we find that (\ref{a}) reduces to the configuration (\ref{l}).
In this case, the boundary condition also changes. One of walls goes
away to infinity and the rest of the wall becomes an elementary wall
 labeled by $\la 2 \leftarrow 3\ra$.

\begin{figure}[h!]
\begin{center}
 $\begin{array}{ccc}
   \epsfxsize=5cm
   \epsfbox{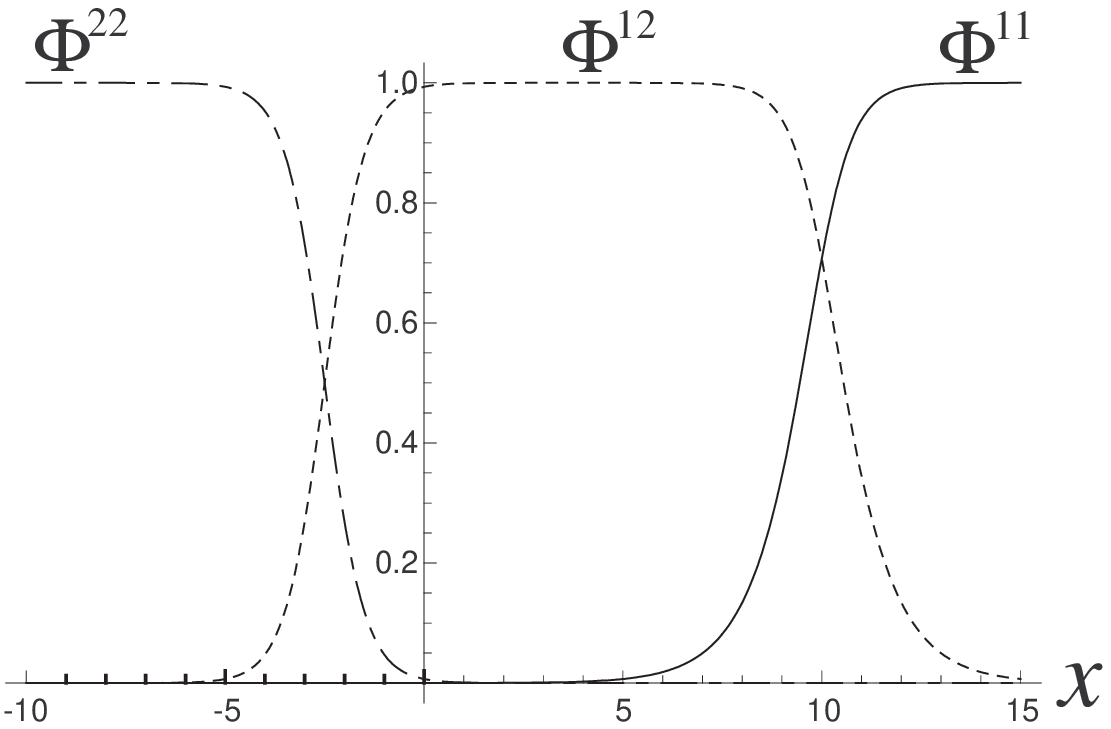}
  &
  \epsfxsize=5cm
   \epsfbox{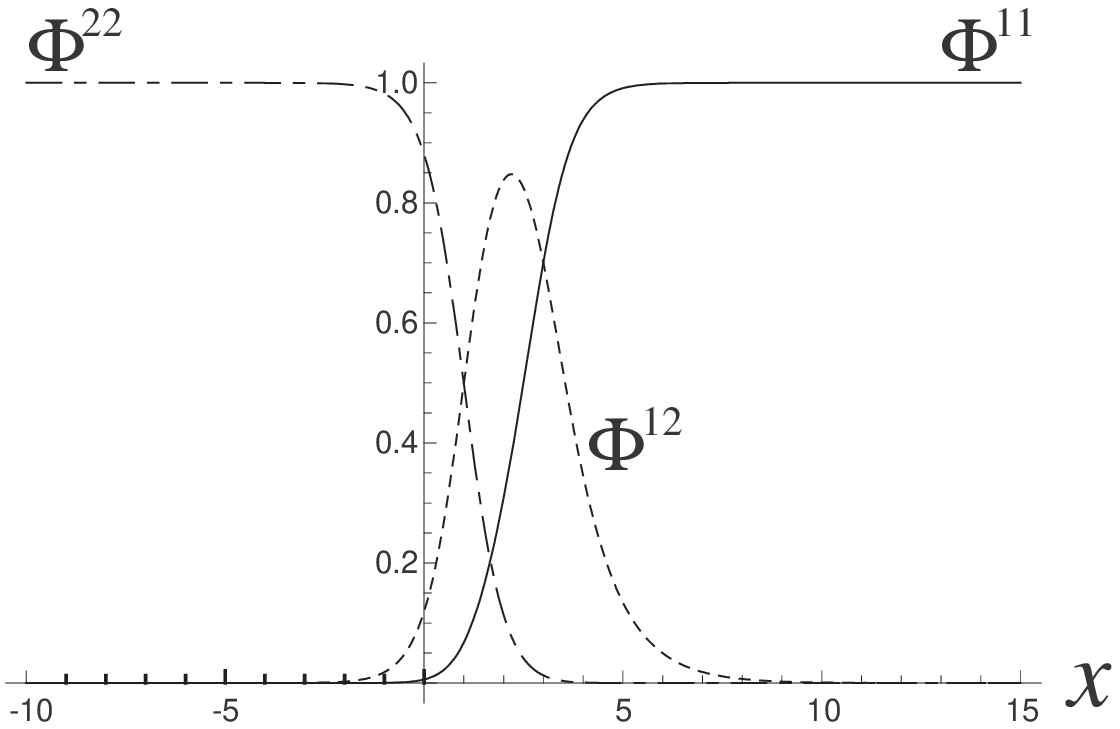}
  &
  \epsfxsize=5cm
   \epsfbox{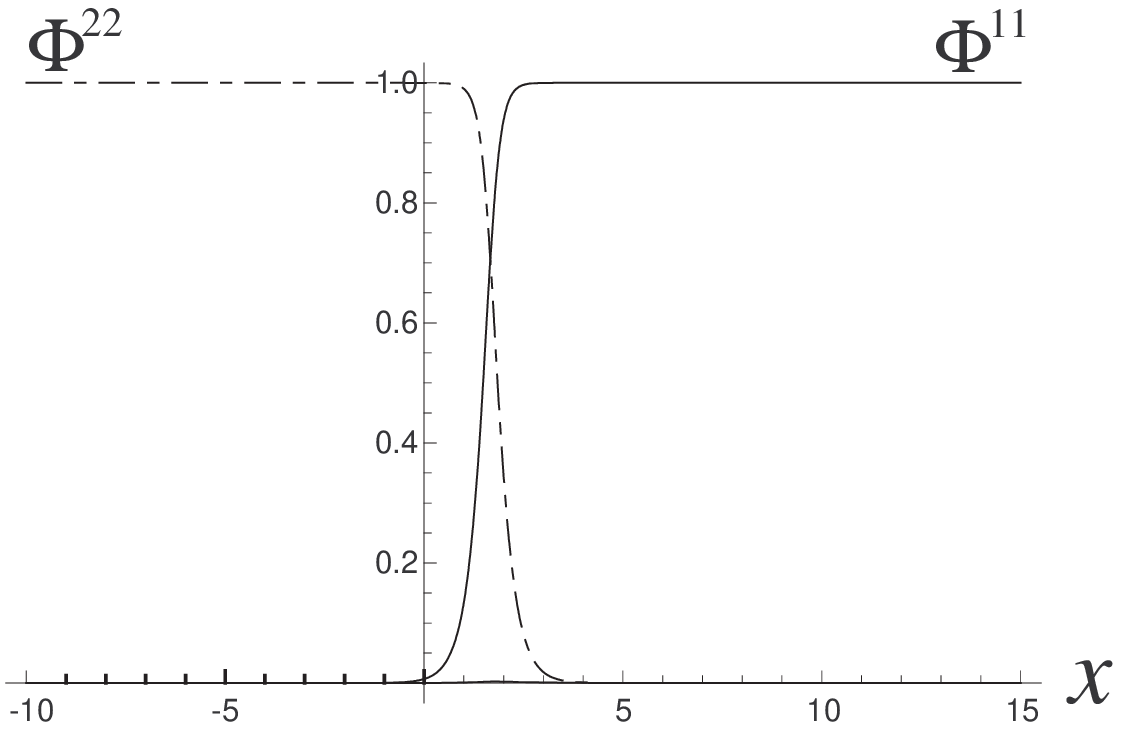}
  \\
  \epsfxsize=5cm
   \epsfbox{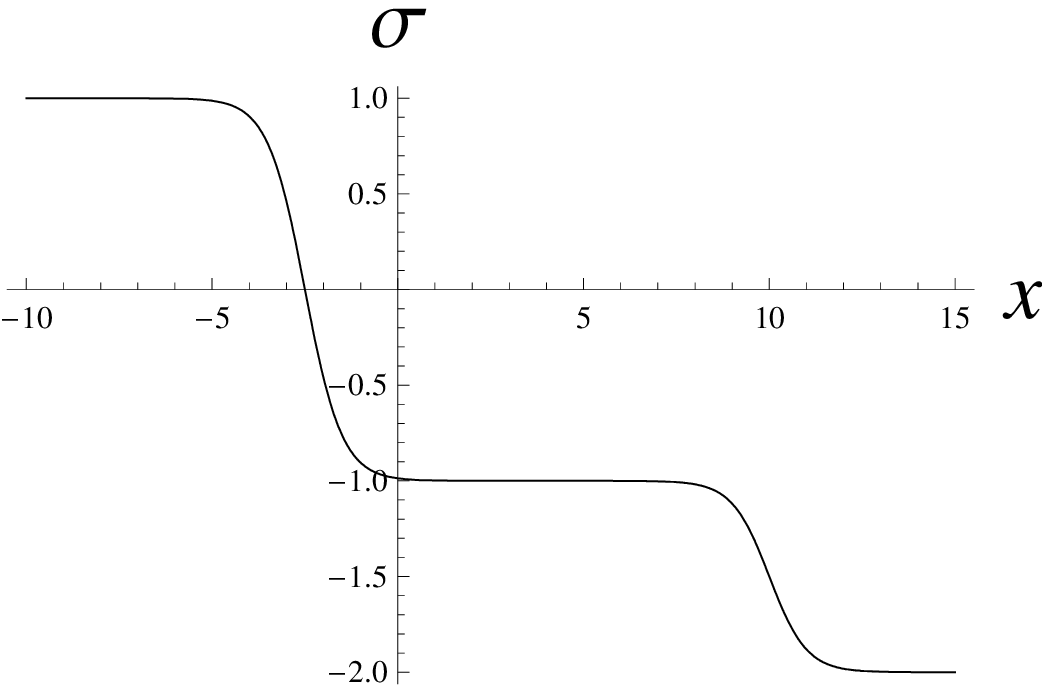}
  &
  \epsfxsize=5cm
   \epsfbox{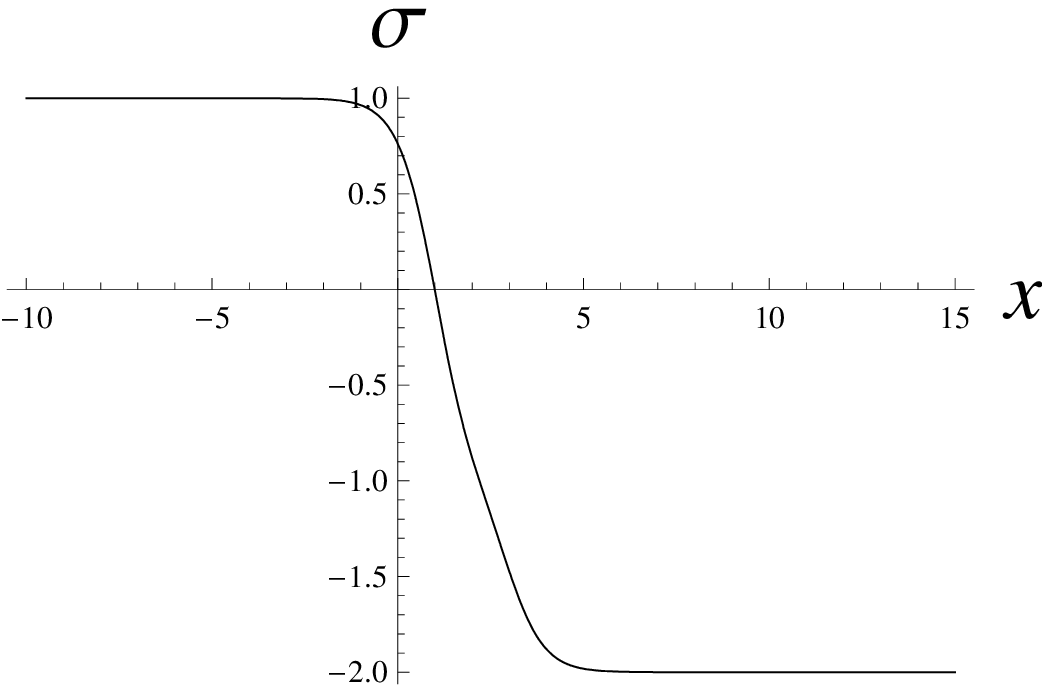}
  &
  \epsfxsize=5cm
   \epsfbox{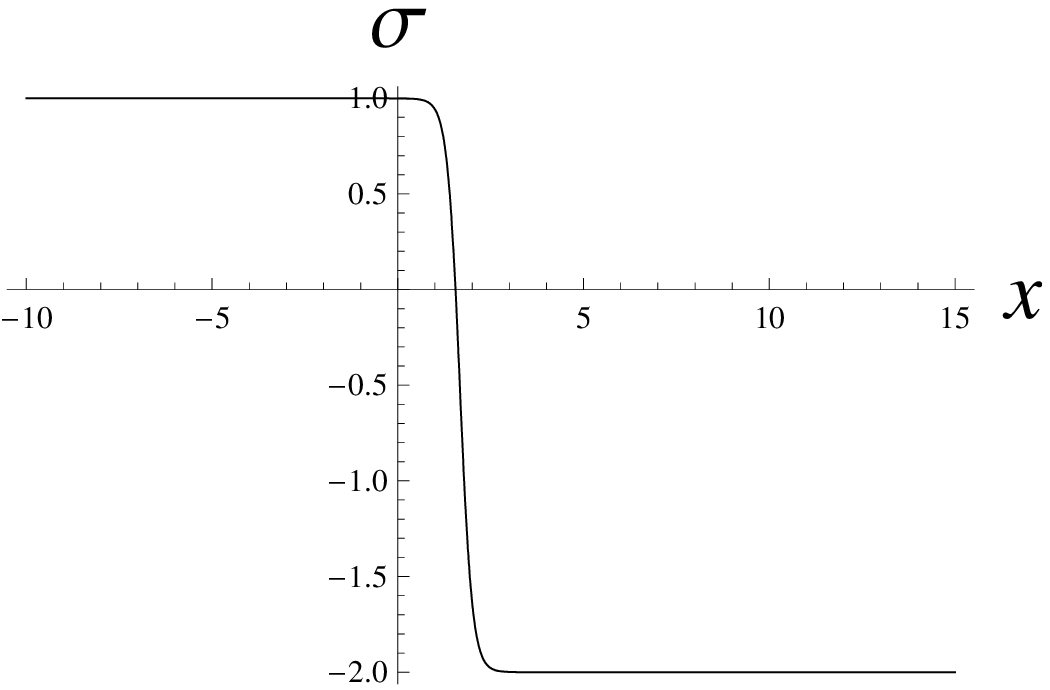}
  \end{array}
 $
\caption{Plots of $\Phi$ and $\sigma$ with $m_1=2,~m_2=1$, $r_3=5$
 and some values of $r_2$.
The left figure ($r_2=10$) shows that there are two separated walls.
As $r_2$ decreases, the left wall approaches the right wall
 (middle figure, $r_2=3$) and eventually both walls are compressed
 into a single wall (right figure, $r_2=-4$).
Even if $r_2$ further decreases, a position and a shape of a
 compressed wall do not change any more.}
\label{comp}
\end{center}
\end{figure}

Finally let us take the limit of $r_2\rightarrow -\infty$.
The configuration (\ref{a}) becomes
\ba
H_{0 \la 1 \leftarrow 3 \ra}=\pmatrix {1 & 0 \cr 0 & e^{r_3}},\quad H_0^5=0. \label{e}
\ea
In this limit, the boundary condition does not change unlike the
 previous two cases. This transition means that two walls approach
 each other and are compressed to a single wall. Therefore, the configuration
 (\ref{e}) exhibits a compressed wall. According to the definition
 below (\ref{comp1}), this is a compressed wall
 of level one. This situation is in contrast to the penetrable walls
 that appeared in the $Q^2$ case.
In that case, a pair of walls just commutes and does not form a
compressed wall. A compressed wall here is formed by two elementary
walls $(\ref{l})$ and $(\ref{a1})$. In general, if a single wall is
composed as compression of $n+1$ elementary walls
 it is called a compressed wall of level $n$.
In Fig. \ref{comp}, we show an example of a generation of a
compressed wall from
 a double wall configuration, according to this explanation.

Similarly, it can be shown that the configuration (\ref{b}) yields
 two elementary wall configurations $H_{0 \la 2 \leftarrow 3\ra}$ and $H_{0 \la 3 \leftarrow 4\ra}$
 and one compressed wall configuration $H_{0 \la 2 \leftarrow 4\ra}$ by taking some limit of
 moduli parameters.

Elementary and compressed walls which appear in the above compose other double wall
 configurations (\ref{c}) and (\ref{d}).
For instance, (\ref{c}) is composed of one elementary wall $H_{0 \la 1\leftarrow 2\ra}$ and
 one compressed wall $H_{0 \la 2\leftarrow 4\ra}$.
The configurations (\ref{c}) and (\ref{d}) can be compressed into a single wall (\ref{k}) by
 taking the limit of $r_2\rightarrow -\infty$ and $r_3\rightarrow -\infty$, respectively.
This single wall (\ref{k}) is compression of three elementary walls and therefore
 it is a compressed wall of level two.

Finally, we show a triple wall configuration. The corresponding
 moduli matrix is obtained by taking all components in $H_0^{\a a}$
 to be nonzero values:
\be
H_{0 \la 1\leftarrow 2\leftarrow 3 \leftarrow 4 \ra}
 =\pmatrix {1&e^{r_2}\cr e^{r_4}&e^{r_3} },\quad H_0^5=\sqrt{2}(e^{r_2+r_3}+e^{r_4})^{1/2}, \label{triple}
\ee
where we take one of the factors to be a  unit by using
(\ref{equiv}). This triple wall configuration is composed of three
elementary walls. Repeating the similar analysis as in a double
wall, it is found that (\ref{triple}) can be reduced
 to all the elementary and double walls that have appeared here.

We show the diagrams representing double and triple wall configurations in the middle
 and right figures in Fig. \ref{Q3-summary}, respectively.
There are four double walls in which two double walls are composed of two elementary walls
 and the others are composed of
 one elementary wall and one compressed wall of level one.
There is only one triple wall composed by three elementary walls.

%
%
\subsection{$N=4$ case}\label{sec;Q4}
In this case, there are $2[4/2+1]=6$ discrete vacua.
As was discussed in Section \ref{sec;massiveKahler},
  this result is consistent with one of the massive HK NLSM on $T^*G_{4,2}$.
It has been shown that
 in the latter model
 there exist
 six elementary walls, five double walls, two triple walls and one
 quadruple wall composed of only elementary walls \cite{INOS1}.
In this subsection, we will show that our result is consistent with theirs.

In the present case, the moduli matrix is written as a 2 times 3
 matrix $H_0^{\a a}$ ($\a=1,2$, $a=1,2,3$).
The explicit representation of vacua is given by
\begin{eqnarray}
&
 H_{0\la 1 \ra}=\left(
  \begin{array}{ccc}
   1 & 0 & 0 \\
   0 & 0 & 0
  \end{array}
 \right),~~
 H_{0\la 2 \ra}=\left(
  \begin{array}{ccc}
   0 & 1 & 0 \\
   0 & 0 & 0
  \end{array}
 \right),~~
 H_{0\la 3 \ra}=\left(
  \begin{array}{ccc}
   0 & 0 & 1 \\
   0 & 0 & 0
  \end{array}
 \right),
& \\
&
 H_{0\la 4 \ra}=\left(
  \begin{array}{ccc}
   0 & 0 & 0 \\
   0 & 0 & 1
  \end{array}
 \right),~~
 H_{0\la 5 \ra}=\left(
  \begin{array}{ccc}
   0 & 0 & 0 \\
   0 & 1 & 0
  \end{array}
 \right),~~
 H_{0\la 6 \ra}=\left(
  \begin{array}{ccc}
   0 & 0 & 0 \\
   1 & 0 & 0
  \end{array}
 \right).
&
\end{eqnarray}
Six elementary single walls connect two of the vacua as listed
below.
\begin{eqnarray}
& H_{0\la 1 \leftarrow 2 \ra}=\left(
  \begin{array}{ccc}
   1 & e^r & 0 \\
   0 & 0 & 0
  \end{array}
 \right),~~
H_{0\la 2 \leftarrow 3 \ra}=\left(
  \begin{array}{ccc}
   0 & 1 & e^r \\
   0 & 0 & 0
  \end{array}
 \right),~~
 H_{0\la 2 \leftarrow 4 \ra}=\left(
  \begin{array}{ccc}
   0 & 1 & 0 \\
   0 & 0 & e^r
  \end{array}
 \right),
& \\
& H_{0\la 3 \leftarrow 5 \ra}=\left(
  \begin{array}{ccc}
   0 & 0 & e^r \\
   0 & 1 & 0
  \end{array}
 \right),~~
H_{0\la 4 \leftarrow 5 \ra}=\left(
  \begin{array}{ccc}
   0 & 0 & 0 \\
   0 & e^r & 1
  \end{array}
 \right),~~
H_{0\la 5 \leftarrow 6 \ra}=\left(
  \begin{array}{ccc}
   0 & 0 & 0 \\
   e^r & 1 & 0
  \end{array}
 \right).
&
\end{eqnarray}
We show the diagram of possible single wall configurations including
 elementary
 and compressed single walls in Fig. \ref{fig;Q4-single}
 in Appendix \ref{appendix-A}.

\begin{figure}[h!]
\begin{center}
 $
 \begin{array}{ccc}
  \epsfxsize=5cm
   \epsfbox{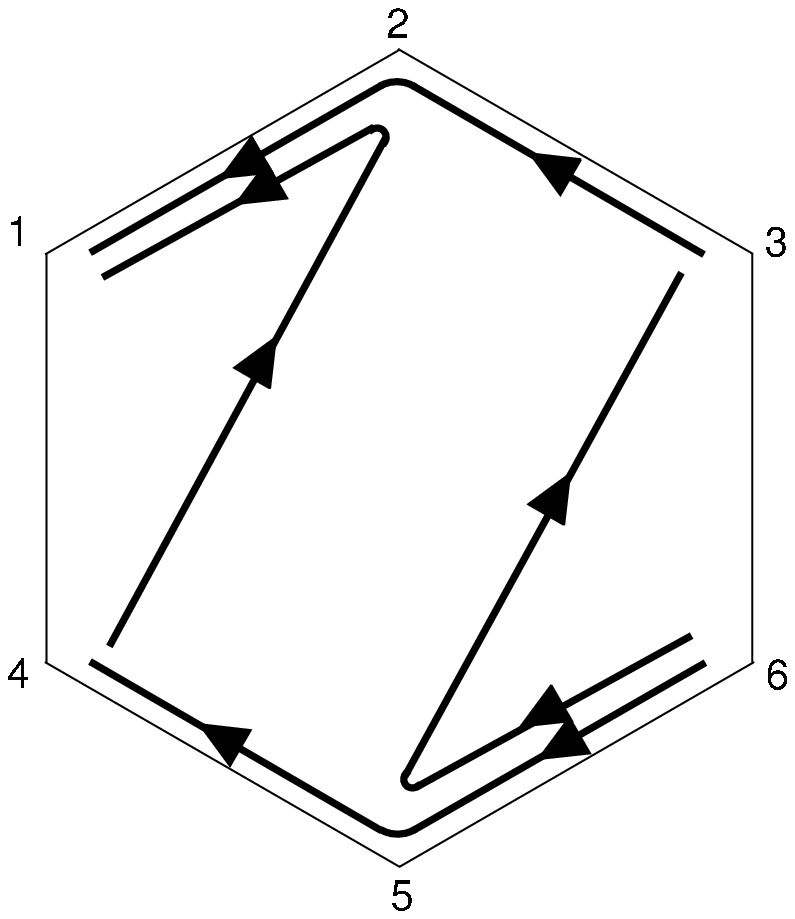}
  & &
    \epsfxsize=5cm
   \epsfbox{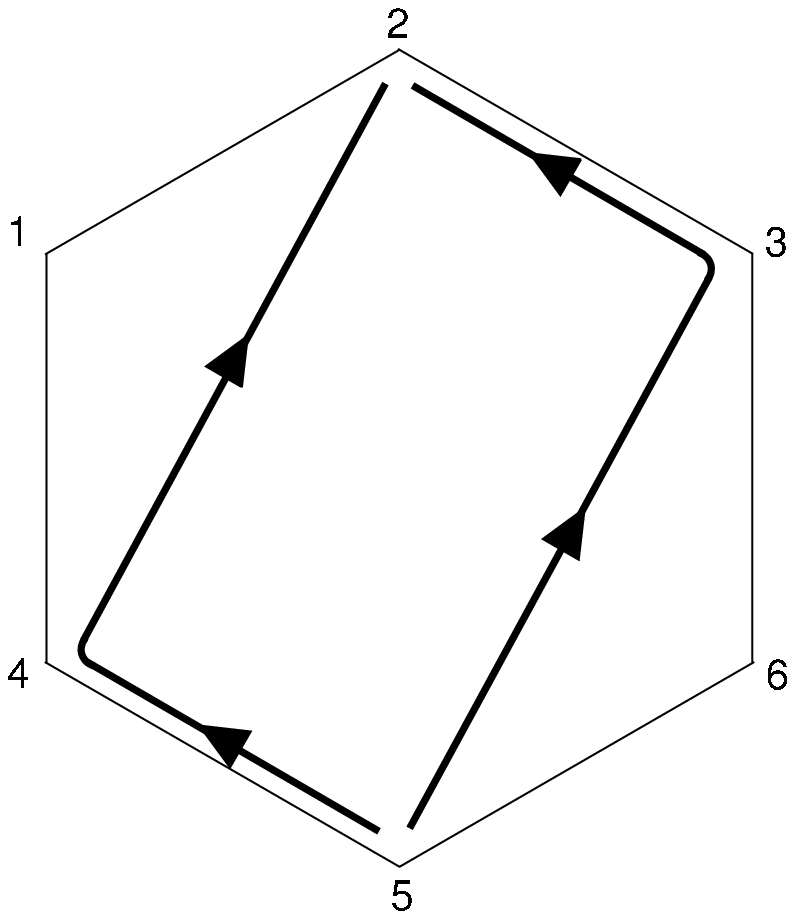}
  \end{array}
 $
\caption{The left figure shows double walls satisfying the constraint (\ref{const-Q4}) trivially.
The right figure shows a double wall $H_{0\la 2\leftarrow 5\ra}$
 connecting either the vacuum labeled by $\la 3\ra$ or the vacuum labeled by $\la 4 \ra$.
Its explicit interpolation of vacua of
 $H_{0\la 2\leftarrow 5\ra}$ is represented by
 either $H_{0\la 2\leftarrow 3\leftarrow 5\ra}$ or
 $H_{0\la 2\leftarrow 4\leftarrow 5\ra}$.
} \label{fig;Q4-double}
\end{center}
\end{figure}

Next we consider double wall configurations.
There are five double walls composed of only elementary walls.
Four of them trivially satisfy the constraints (\ref{const-Q4}).
The moduli matrices are written as
\begin{eqnarray}
& H_{0\la 1 \leftarrow 2 \leftarrow 3\ra}=\left(
  \begin{array}{ccc}
   1 & e^{r_2} & e^{r_3} \\
   0 & 0 & 0
  \end{array}
 \right),~~
H_{0\la 1 \leftarrow 2 \leftarrow 4 \ra}=\left(
  \begin{array}{ccc}
   1 & e^{r_2} & 0 \\
   0 & 0 & e^{r_4}
  \end{array}
 \right), \label{d1}
& \\
&H_{0\la 3 \leftarrow 5 \leftarrow 6 \ra}=\left(
  \begin{array}{ccc}
   0 & 0 & 1 \\
   e^{r_6} & e^{r_5} & 0
  \end{array}
 \right),~~
H_{0\la 4 \leftarrow 5 \leftarrow 6 \ra}=\left(
  \begin{array}{ccc}
   0 & 0 & 0 \\
   e^{r_6} & e^{r_5} & 1
  \end{array}
 \right).
&
\end{eqnarray}
The other double wall is given by
\begin{eqnarray}
H_{0\la 2 \leftarrow 5 \ra}=\left(
  \begin{array}{ccc}
   0 & 1 & e^{r_3} \\
   0 & e^{r_5} & e^{r_4}
  \end{array}
 \right),\label{Q4-ele-dou}
\end{eqnarray}
with the constraint
\begin{eqnarray}
e^{r_3+r_4}+e^{r_5}=0. \label{Q4-const}
\end{eqnarray}
This is exactly the same equation as (\ref{c-Q2}).
Repeating the same analysis below (\ref{c-Q2}), for the
 configuration (\ref{Q4-ele-dou}) we find two possible parameter regions
\begin{eqnarray}
&& x_2=x_4 \gg x_3 \gg x_5, \label{q4;c-1}\\
&& x_3 \gg x_2=x_4 \gg x_5, \label{q4;c-2}
\end{eqnarray}
where $x_l={\rm Re}(X_l)$ is defined in (\ref{position}) with
 $r_2=0$. The former parametrization excludes the vacuum labeled
 by $\la 4\ra$ whereas the
 latter one excludes
 the vacuum $\la 3\ra$ in the configuration.
The moduli matrix (\ref{Q4-ele-dou}) with the constraint
(\ref{Q4-const})
 therefore connects
 vacua labeled by $\la 2 \ra$, $\la 3\ra~(\la 4 \ra)$ and $\la 5 \ra$
 under the former (latter) parameter region.
A pair of walls in this configuration is a penetrable wall, as we discussed
 in Section \ref{sec;N=2}.
In this case,
 the transition from $H_{0\la 2\leftarrow 3\leftarrow 5\ra}$ to
 $H_{0\la 2\leftarrow 4\leftarrow 5\ra}$ (and vice versa) occurs
 as the positions of two walls exchange.
All the double walls composed of
 only elementary walls for (\ref{q4;c-1}) and (\ref{q4;c-2}) are depicted in Fig. \ref{fig;Q4-double}.
%

\begin{figure}[h!]
\begin{center}
 $\begin{array}{ccc}
  \epsfxsize=5cm
   \epsfbox{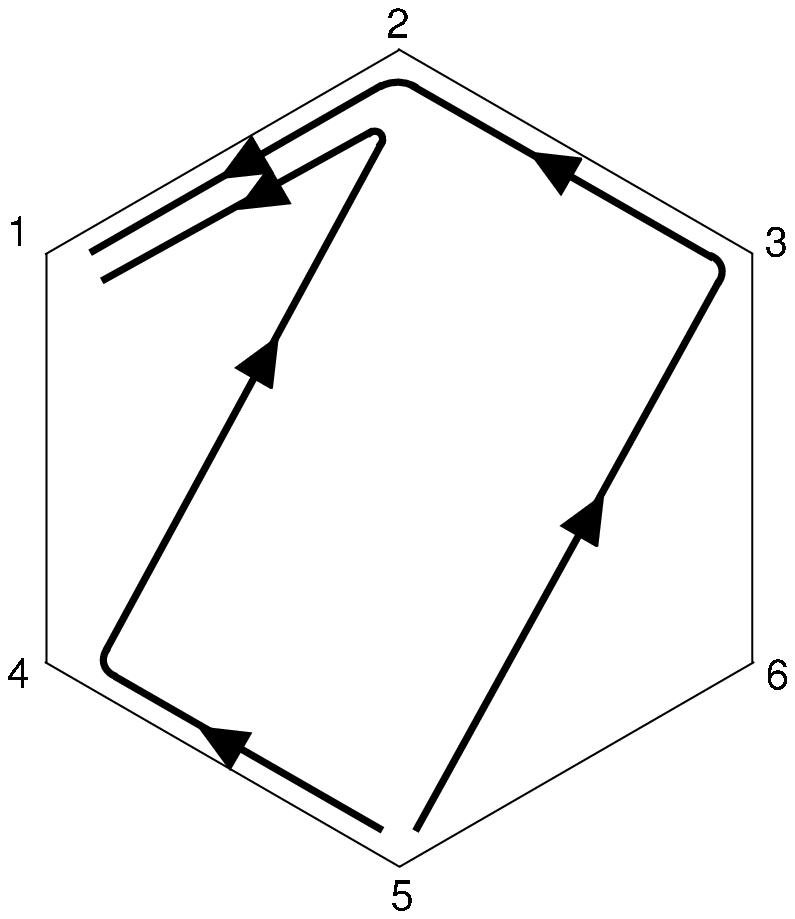}
& &
  \epsfxsize=5cm
   \epsfbox{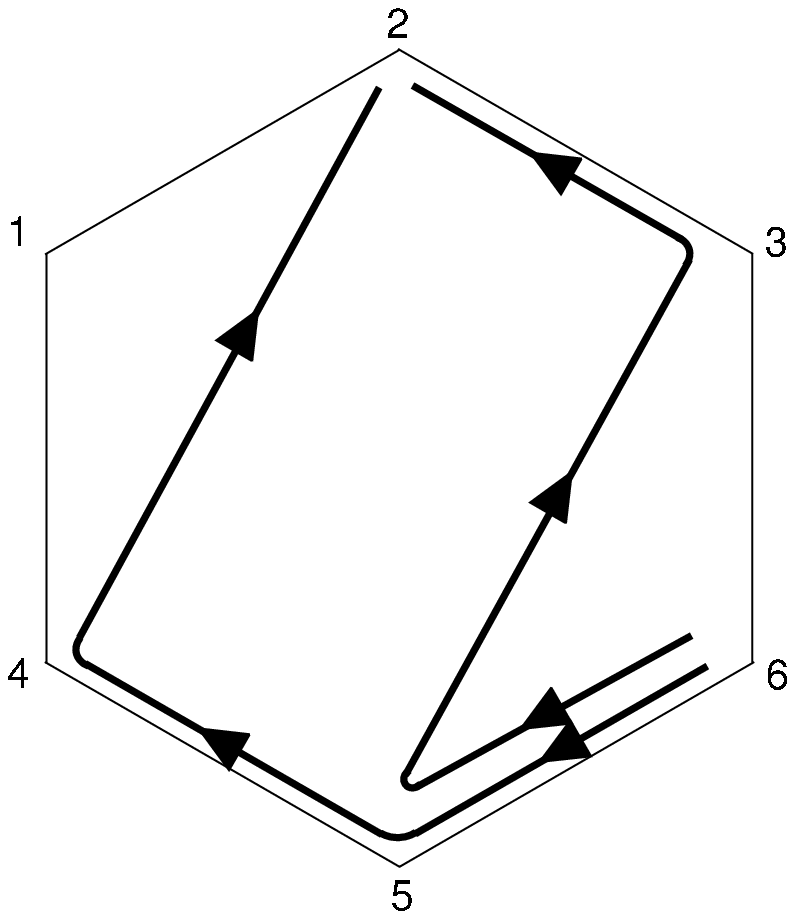}
 \end{array}$
 \caption{The left figure describes a triple wall $H_{0\la 1 \leftarrow 5\ra}$,
whose explicit interpolation of vacua is represented by
 either $H_{0\la 1 \leftarrow 2 \leftarrow 3\leftarrow 5\ra}$ or
 $H_{0\la 1 \leftarrow 2 \leftarrow 4 \leftarrow 5\ra}$.
The right figure describes a triple wall
 $H_{0\la 2 \leftarrow 6 \ra}$.
Explicit interpolation of vacua is represented by
 either
 $H_{0\la 2 \leftarrow 3 \leftarrow 5 \leftarrow 6\ra}$
 or $H_{0\la 2 \leftarrow 4 \leftarrow 5 \leftarrow 6\ra}$.}\label{fig;Q4-triple}
\end{center}
\end{figure}

There are two triple walls composed of only elementary walls.
The moduli matrices are
\begin{eqnarray}
H_{0\la 1 \leftarrow 5\ra}=\left(
  \begin{array}{ccc}
   1 & e^{r_2} & e^{r_3} \\
   0 & e^{r_5} & e^{r_4}
  \end{array}
 \right),\quad e^{r_2+r_5}+e^{r_3+r_4}=0, \label{h15}
\end{eqnarray}
and
\begin{eqnarray}
H_{0\la 2 \leftarrow 6 \ra}=\left(
  \begin{array}{ccc}
   0 & 1 & e^{r_3} \\
   e^{r_6} & e^{r_5} & e^{r_4}
  \end{array}
 \right),\quad e^{r_3+r_4}+e^{r_5}=0. \label{h26}
\end{eqnarray}
Again constraints appeared here are the same form as (\ref{Q4-const}).
Repeating the same discussion, it is found that
 the triple wall $H_{0\la 1 \leftarrow 5\ra}$ connects
 the vacua
 labeled by $\la 1 \ra$, $\la 2 \ra$, $\la 3 \ra~(\la 4 \ra)$ and $\la 5 \ra$,
 and
 $H_{0\la 2 \leftarrow 6 \ra}$ connects
 the vacua $\la 2 \ra$, $\la 3 \ra~(\la 4 \ra)$,
 $\la 5 \ra$ and $\la 6 \ra$.
These two configurations include penetrable walls.
They are single walls labeled by
 $\la 2\leftarrow 3 \ra~(\la 2\leftarrow 4 \ra)$
 and $\la 3\leftarrow 5 \ra~(\la 4\leftarrow 5 \ra)$ for both configurations.
The diagrams of the triple walls are given in Fig. \ref{fig;Q4-triple}.
%

The moduli matrix for a quadruple wall which is composed of the maximal number of
 single walls is given by
\begin{eqnarray}
H_{0\la 1 \leftarrow 6 \ra}=\left(
  \begin{array}{ccc}
   1 & e^{r_2} & e^{r_3} \\
   e^{r_6} & e^{r_5} & e^{r_4}
  \end{array}
 \right),\label{Q4-4wall}
\end{eqnarray}
with the constraint
\begin{eqnarray}
e^{r_2+r_5}+e^{r_3+r_4}+e^{r_6}=0. \label{Q4-4wall-const}
\end{eqnarray}
It interpolates five among six vacua
 and it is therefore composed of
 elementary walls only.
We will see this in detail below.

\begin{figure}[h!]
\begin{center}
  \epsfxsize=5cm
   \epsfbox{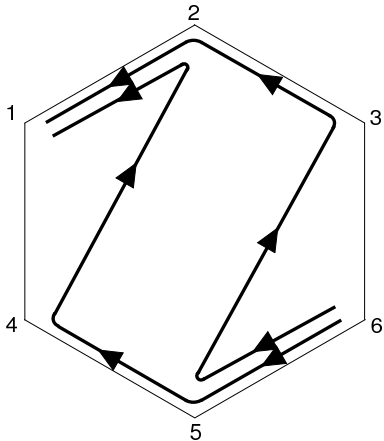}
 \caption{The quadruple wall $H_{0\la 1 \leftarrow 6 \ra}$.
  Explicit interpolation of vacua
  is represented by either
  $H_{0\la 1 \leftarrow 2 \leftarrow 3 \leftarrow 5 \leftarrow 6 \ra}$ or
 $H_{0\la 1 \leftarrow 2 \leftarrow 4 \leftarrow 5 \leftarrow 6 \ra}$.}\label{fig;Q4-quadruple}
\end{center}
\end{figure}

First of all let us solve the constraint.
The constraint can be solved in the fashion analogous to the previous one.
Given the transformation (\ref{shift}) and the
 relation (\ref{position}) with $r_1=0$ the constraint (\ref{Q4-4wall-const}) is
 written in the form
\begin{eqnarray}
e^{(m_4-m_5)X_4+(m_5-m_6)X_5}+e^{(m_1-m_2)X_1+(m_4-m_5)X_4}+e^{(m_1-m_2)X_1+(m_2-m_3)X_2}=0,\label{Q4-4wall-const;a}
\end{eqnarray}
where $m_4=-m_3$, $m_5=-m_2$ and $m_6=-m_1$.
This can be rewritten as
\begin{eqnarray}
e^{-(m_4-m_5)X_4}=-e^{-(m_2-m_3)X_2}\left[1+e^{-(m_1-m_2)(X_1-X_5)}\right].\label{Q4-4wall-const;b}
\end{eqnarray}
Assuming that $x_1 \gg x_5$, the constraint (\ref{Q4-4wall-const;b})
 can be approximately solved by
\begin{eqnarray}
X_4 \sim X_2+\frac{i(2n+1)\pi}{m_2-m_3},\quad n\in Z.
\end{eqnarray}
This tells us that
\begin{eqnarray}
x_2 \sim x_4. \label{q4;1gg5}
\end{eqnarray}
Taking into account this, we can choose the following parameter regions:
\begin{eqnarray}
&& x_1\gg x_2 \sim x_4 \gg x_3 \gg x_5 \gg x_6, \label{q4;c-3}\\
&& x_1\gg x_3 \gg x_2 \sim x_4 \gg x_5 \gg x_6. \label{q4;c-4}
\end{eqnarray}
The above regions include the same parameter regions with (\ref{q4;c-1}) and
 (\ref{q4;c-2}), respectively.
It is therefore found that the moduli matrix (\ref{Q4-4wall})
 describes a quadruple wall connecting all the vacua except
 either the vacua $\la 3 \ra$ or $\la 4 \ra$.
The quadruple wall includes penetrable walls.
They are labeled by $\la 2 \leftarrow 3\ra$ and $\la 3\leftarrow 5\ra$ for the region (\ref{q4;c-3}),
 and $\la 2 \leftarrow 4 \ra$ and $\la 4 \leftarrow 5 \ra$ for the region (\ref{q4;c-4}).
The diagram of the configurations
 is illustrated in Fig. \ref{fig;Q4-quadruple}.
For the configuration
 (\ref{Q4-4wall}), there are other possible parameter choices.
They are shown in Appendix \ref{appendix-B}.
%
%

Moduli matrices for compressed single walls and multiwalls composed
 of compressed walls are summarized in Appendix \ref{appendix-A}.

%
%
\section{Conclusion}
We have investigated the vacuum structure and the exact BPS domain
wall solutions
 in the massive K\"ahler NLSM on the complex quadric surface $Q^N$.
This massive model has been obtained from the massless K\"ahler NLSM on $Q^N$ in
 4-dimensional space-time by the Scherk-Schwarz dimensional reduction.
The dimensional reduction gives rise to a nontrivial scalar
potential which
 includes mass terms characterized by the Cartan matrix of $SO(N+2)$.
We have assumed generic mass parameters and
 have found that the theory has $2[N/2+1]$ discrete vacua.
The exact BPS multiwall solutions have been derived based on the moduli matrix
 approach developed in \cite{INOS1,INOS2}, especially for the $N\le 4$ case.
We have also found that the moduli space of the wall solutions is the complex
 quadric surface.

We have discussed the consistency of our result with ones
 previously studied. As mentioned in the Introduction and Section
 \ref{sec;massiveKahler},
 when considering vacua and walls in the massive HK NLSM on
 $T^*{\bf C}P^1$ and $T^*G_{N_F,N_C}$, the cotangent part is irrelevant.
We can respect the results in these models as ones in the massive K\"ahler
 NLSMs on ${\bf C}P^1$ and $G_{N_F,N_C}$.
For instance, the massive HK NLSM on $T^*{\bf C}P^1$ gives two discrete vacua
 and one domain wall solution interpolating them.
Vacua and wall solutions are only described by the ${\bf C}P^1$ part
 of the model. Since ${\bf C}P^1$ is isomorphic to $Q^1$, the same
 result should be obtained
 in the $Q^1$ case.
Actually we have found the same number of discrete vacua and domain
 wall solutions.
Similarly, we have checked the consistency for
 the $Q^4$ case.
This is isomorphic to $G_{4,2}$ and the results of the $Q^4$
 case should be
 the same with one of the massive HK NLSM on $T^*G_{4,2}$.
The latter possesses six discrete vacua and the BPS wall solutions consisting of
 six elementary single walls, five double walls, two triple walls and one quadruple
 wall composed of only elementary walls.
There also exist single compressed walls and multiwalls including
 compressed walls. Our wall solutions in the $Q^4$ case completely
 coincide with these results.

The results of the $Q^2$ and $Q^3$ cases are completely new although
 there is isomorphism $Q^2\sim {\bf C}P^1\times {\bf C}P^1$ and
 $Q^3\sim Sp(2)/U(2)$. It can be
 guessed that our $Q^2$ model has four discrete vacua since it consists of two
 ${\bf C}P^1$ sectors.
This has been justified in our analysis.
 The $Q^3$ model has a richer structure than the $Q^2$ model
 despite the fact that the number of the vacua is the same in both
 cases. The reason is that the constraint (\ref{const-Q4}) is
 different. The former includes the scalar $H_0^{N+2}$ in
(\ref{const-Q4}) while the latter does not. When it exists in
(\ref{const-Q4}), any choice of moduli parameters in $H_0^{\a a}$ is
 possible, since such a choice is compensated by the scalar $H_0^{N+2}$ through
 (\ref{const-Q4}).
The same structure is repeated generally for $Q^{2n}$ and $Q^{2n+1}~(n \ge 1)$.
Though both models have the same numbers of discrete vacua $2(n+1)$,
 since the latter includes $H_0^{2n+1}$, there are more abundant types of domain wall
 solutions than the $Q^{2n}$ case.
Therefore, the $Q^5$ model has more kinds of wall solutions than the
 $Q^4$ case studied here. Deriving wall solutions in the
 $N=5$ case is very similar to the $N=3$ case and
 therefore we do not repeat these here.

We believe that our results obtained in this paper are the same as
 those of the massive HK NLSM on $T^*Q^N$ for the same reason
 mentioned above. One way to see is to construct a quotient action of
 this model
 and to derive wall solutions with the help of the moduli matrix approach.
However, as we wrote in the Introduction,
 construction of a quotient action of this model is difficult though
 its massless version without using a Lagrange multiplier has been constructed
 \cite{AKL1}.
The latter model can be extended into a massive version by using the formulation
 in \cite{Kuzenko}.
In this formulation, we cannot directly use the moduli matrix approach since it is
 not written as a gauge theory.
It would be interesting to investigate the vacuum structure and wall
solutions in
 this model to check the consistency as a future work.

%
%
\vspace{8mm}
\noindent{\Large \bf Acknowledgments}\\
We would like to thank Minoru Eto, Hiroaki Nakajima and Muneto Nitta  for discussion.
The work of M.A. is supported by the Science Research Center Program
of the Korea Science and Engineering Foundation through the Center for
 Quantum Spacetime (CQUeST) of Sogang University with grant number
 R11-2005-021.
S.L. would like to thank CQUeST for hospitality during this work.

%
%
\vspace{8mm}
\noindent{\Large \bf Appendix}
\appendix
\setcounter{equation}{0}
\section{Compressed walls for $Q^4$}\label{appendix-A}
We discuss single walls and multiwalls composed of compressed walls
for the $N=4$ case.

\begin{figure}[h!]
\begin{center}
  \epsfxsize=5cm
   \epsfbox{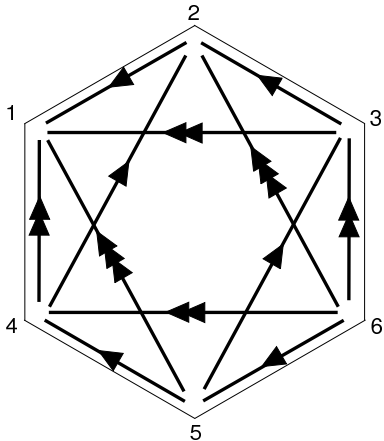}
 \caption{Arrows with one arrowhead denote elementary walls, which are discussed in Section \ref{sec;Q4}.
 Arrows with two arrowheads denote
 compressed single walls of level one. Arrows with three arrowheads denote
 compressed single walls of level two. All the arrows point from
 $x=-\infty$ to $x=\infty$.
 }\label{fig;Q4-single}
\end{center}
\end{figure}

There are four compressed single walls of level one:
\begin{eqnarray}
& H_{0\la 1 \leftarrow 4\ra}=\left(
  \begin{array}{ccc}
   1 & 0 & 0 \\
   0 & 0 & e^r
  \end{array}
 \right),~~
H_{0\la 1 \leftarrow 3 \ra}=\left(
  \begin{array}{ccc}
   1 & 0 & e^r \\
   0 & 0 & 0
  \end{array}
 \right),
& \nonumber\\
&H_{0\la 4 \leftarrow 6 \ra}=\left(
  \begin{array}{ccc}
   0 & 0 & 0 \\
   e^r & 0 & 1
  \end{array}
 \right),~~
H_{0\la 3 \leftarrow 6 \ra}=\left(
  \begin{array}{ccc}
   0 & 0 & 1 \\
   e^r & 0 & 0
  \end{array}
 \right).
&
\end{eqnarray}
There are two compressed single walls of level two:
\begin{eqnarray}
&H_{0\la 1 \leftarrow 5 \ra}=\left(
  \begin{array}{ccc}
   1 & 0 & 0 \\
   0 & e^r & 0
  \end{array}
 \right),~~
H_{0\la 2 \leftarrow 6 \ra}=\left(
  \begin{array}{ccc}
   0 & 1 & 0 \\
   e^r & 0 & 0
  \end{array}
 \right).
&
\end{eqnarray}
The diagram of the above configurations are depicted in Fig. \ref{fig;Q4-single}.
The compressed single walls can be constructed from the multiwalls as discussed
 in Section \ref{sec;N=2}.
It is also read off from this in the diagram in Fig.
\ref{fig;Q4-single}. For instance, it is easy to see that the
compressed wall $H_{0\la 1\leftarrow 4\ra}$
 can be obtained as a compression of the single walls $\la 1 \leftarrow 2\ra$ and
 $\la 2\leftarrow 4\ra$ in $H_{0\la 1\leftarrow 2 \leftarrow 4\ra}$ in (\ref{d1}).

There exist four double walls composed of compressed walls of level
one:
\begin{eqnarray}
& H_{0\la 1 \leftarrow 3 \leftarrow 5\ra}=\left(
  \begin{array}{ccc}
   1 & 0 & e^{r_3} \\
   0 & e^{r_5} & 0
  \end{array}
 \right),~~
H_{0\la 1 \leftarrow 4 \leftarrow 5 \ra}=\left(
  \begin{array}{ccc}
   1 & 0 & 0 \\
   0 & e^{r_5} & e^{r_4}
  \end{array}
 \right),
& \nonumber\\
&H_{0\la 2 \leftarrow 4 \leftarrow 6 \ra}=\left(
  \begin{array}{ccc}
   0 & 1 & 0 \\
   e^{r_6} & 0 & e^{r_4}
  \end{array}
 \right),~~
H_{0\la 2 \leftarrow 3 \leftarrow 6 \ra}=\left(
  \begin{array}{ccc}
   0 & 1 & e^{r_3}  \\
   e^{r_6} & 0 & 0
  \end{array}
 \right).
&
\end{eqnarray}
A double wall
\begin{eqnarray}
H_{0\la 1 \leftarrow 6 \ra}= \left(\begin{array}{ccc}
   1 & e^{r_2} & 0 \\
   e^{r_6} & e^{r_5} & 0
  \end{array}
 \right),~~
 ~e^{r_2+r_5}+e^{r_6}=0,
\end{eqnarray}
is composed of an elementary wall and a compressed wall of level
 two.
Its explicit interpolation of vacua is represented by
 either $H_{0\la 1 \leftarrow 2 \leftarrow 6 \ra}$
 or $H_{0\la 1 \leftarrow 5 \leftarrow 6 \ra}$.
A double wall
\begin{eqnarray}
H_{0\la 1 \leftarrow 6 \ra}= \left(\begin{array}{ccc}
   1 & 0 & e^{r_3} \\
   e^{r_6} & 0 & e^{r_4}
  \end{array}
 \right),~~
 ~e^{r_3+r_4}+e^{r_6}=0,
\end{eqnarray}
is composed of two compressed walls of level two.
Its explicit interpolation of vacua is represented by
 either $H_{0\la 1 \leftarrow 3 \leftarrow 6 \ra}$ or $H_{0\la 1
 \leftarrow 4 \leftarrow 6 \ra}$.

There are two triple walls composed of two elementary walls and a
 compressed wall of level one:
\begin{eqnarray}
H_{0\la 1 \leftarrow 6 \ra}= \left(\begin{array}{ccc}
   1 & e^{r_2} & e^{r_3} \\
   e^{r_6} & 0 & e^{r_4}
  \end{array}
 \right),~~
H_{0\la 1 \leftarrow 6 \ra}= \left(\begin{array}{ccc}
   1 & 0 & e^{r_3} \\
   e^{r_6} & e^{r_5} & e^{r_4}
  \end{array}
 \right).
\end{eqnarray}
Both are constrained by $e^{r_3+r_4}+e^{r_6}=0$.
Explicit interpolation of vacua for the first one is represented by
 either $H_{0\la 1 \leftarrow 2 \leftarrow 3 \leftarrow 6 \ra}$
 or $H_{0\la 1 \leftarrow 2 \leftarrow 4 \leftarrow 6 \ra}$.
Similarly for the second one it is represented by
 either $H_{0\la 1 \leftarrow 3 \leftarrow 5 \leftarrow 6 \ra}$
 or $H_{0\la 1 \leftarrow 4 \leftarrow 5 \leftarrow 6 \ra}$.

\section{Parameter regions for the configuration (\ref{Q4-4wall})}\label{appendix-B}
In this appendix, we show another choice for parameter regions for the
 configuration (\ref{Q4-4wall}).
Assuming that $x_1 \ll x_5$,
 the constraint (\ref{Q4-4wall-const;b}) leads to $x_2 \ll x_4$.
Considering this, possible choices of parameters are in this case, for instance,
\begin{eqnarray}
x_3 \gg x_4 \gg x_5 \gg x_1 \gg x_2 \gg x_6, \label{q4;d-1}\\
x_4 \gg x_5 \gg x_1 \gg x_2 \gg x_3 \gg x_6. \label{q4;d-2}
\end{eqnarray}
The parameter choice (\ref{q4;d-1}) tells us that a quadruple wall configuration
 interpolates all the vacua except the vacuum $\la 3 \ra$
 since it does not satisfy the condition to lead to this vacuum
 (see the discussion below (\ref{relation1}))
\ba
 x_2 \gg x_0 \gg x_3,
\ea
whereas (\ref{q4;d-2}) excludes the vacuum $\la 4 \ra$ by the similar reason.

The constraint (\ref{Q4-4wall-const;a}) can be also written as
\begin{eqnarray}
e^{(m_5-m_6)X_5}=-e^{(m_1-m_2)X_1}[1+e^{(m_2-m_3)(X_2-X_4)}].\label{Q4-4wall-const;c}
\end{eqnarray}
Assuming that $x_2 \gg x_4$ the constraint (\ref{Q4-4wall-const;c}) leads to
 $x_1 \ll x_5$.
Possible choices for parameter regions are given by
\begin{eqnarray}
x_5 \gg x_1 \gg x_2 \gg x_3 \gg x_4 \gg x_6, \label{q4;e-1} \\
x_2 \gg x_3 \gg x_4 \gg x_5 \gg x_1 \gg x_6. \label{q4;e-2}
\end{eqnarray}
Repeating the same discussion as above, we find that the parameter choices
 (\ref{q4;e-1}) and (\ref{q4;e-2})
 represent configurations of quadruple walls interpolating all the vacua except
 the vacua labeled by $\la 5 \ra$ and $\la 2 \ra$,
 respectively.

Finally, if it is assumed that $x_2 \ll x_4$,
 the constraint (\ref{Q4-4wall-const;c}) gives
 $x_1 \sim x_5$.
Appropriate choices for parametrization are
\begin{eqnarray}
x_3 \gg x_4 \gg x_5\sim x_1 \gg x_2 \gg x_6, \label{q4;f-1}\\
x_4 \gg x_5\sim x_1 \gg x_2 \gg x_3 \gg x_6. \label{q4;f-2}
\end{eqnarray}
In this case,
 the vacua $\la 3 \ra$ and $\la 4 \ra$ are not interpolated
 in the region (\ref{q4;f-1}) and (\ref{q4;f-2}), respectively.

%
%


\begin{thebibliography}{99}
%
\bibitem{WittenOlive}
 E.~Witten and D.~Olive,
  {\it Phys.\ Lett.\ } {\bf B78} (1978) 97.
%
\bibitem{BPS}
 E.~Bogomol'nyi,
  {\it Sov.\ J.\ Nucl.\ Phys.\ } {\bf B24} (1976) 449;
 M.~K.~Prasad and C.~H.~Sommerfield,
  {\it Phys.\ Rev.\ Lett.\ } {\bf 35} (1975) 760.
%
\bibitem{CQR}
  M.~Cvetic, F.~Quevedo and S.~J.~Rey,
   Phys.\ Rev.\ Lett.\  {\bf 67} (1991) 1836;
  M.~Cvetic, S.~Griffies and S.~J.~Rey,
   Nucl.\ Phys.\  B {\bf 381} (1992) 301
   [arXiv:hep-th/9201007];
  M.~Cvetic, S.~Griffies and H.~H.~Soleng,
   Phys.\ Rev.\  D {\bf 48} (1993) 2613
   [arXiv:gr-qc/9306005].
%
\bibitem{AT}
  E.~R.~C.~Abraham and P.~K.~Townsend,
  Phys.\ Lett.\  B {\bf 291} (1992) 85.
%
\bibitem{INOS1}
  Y.~Isozumi, M.~Nitta, K.~Ohashi and N.~Sakai,
   Phys.\ Rev.\ Lett.\  {\bf 93} (2004) 161601
   [arXiv:hep-th/0404198];
  Y.~Isozumi, M.~Nitta, K.~Ohashi and N.~Sakai,
  Phys.\ Rev.\  D {\bf 70} (2004) 125014
  [arXiv:hep-th/0405194].
%
\bibitem{INOS2}
  M.~Eto, Y.~Isozumi, M.~Nitta, K.~Ohashi and N.~Sakai,
  J.\ Phys.\ A  {\bf 39} (2006) R315
  [arXiv:hep-th/0602170].
%
\bibitem{INOS3}
  Y.~Isozumi, M.~Nitta, K.~Ohashi and N.~Sakai,
  Phys.\ Rev.\  D {\bf 71} (2005) 065018
  [arXiv:hep-th/0405129].
%
\bibitem{EINOS1}
  M.~Eto, Y.~Isozumi, M.~Nitta, K.~Ohashi and N.~Sakai,
  Phys.\ Rev.\  D {\bf 72} (2005) 085004
  [arXiv:hep-th/0506135];
  Phys.\ Lett.\  B {\bf 632} (2006) 384
  [arXiv:hep-th/0508241].
%
\bibitem{EINOS2}
  M.~Eto, Y.~Isozumi, M.~Nitta, K.~Ohashi and N.~Sakai,
  Phys.\ Rev.\ Lett.\  {\bf 96} (2006) 161601
  [arXiv:hep-th/0511088];
  M.~Eto, T.~Fujimori, Y.~Isozumi, M.~Nitta, K.~Ohashi, K.~Ohta and N.~Sakai,
  Phys.\ Rev.\  D {\bf 73} (2006) 085008
  [arXiv:hep-th/0601181].
%
\bibitem{EINOS3}
M.~Eto, Y.~Isozumi, M.~Nitta, K.~Ohashi and N.~Sakai,
  Phys.\ Rev.\  D {\bf 72} (2005) 025011
  [arXiv:hep-th/0412048].
%
\bibitem{ENOT}
M.~Eto, M.~Nitta, K.~Ohashi and D.~Tong,
  Phys.\ Rev.\ Lett.\  {\bf 95} (2005) 252003
  [arXiv:hep-th/0508130].
%
\bibitem{ANS}
  M.~Arai, M.~Nitta and N.~Sakai,
  Prog.\ Theor.\ Phys.\  {\bf 113} (2005) 657
  [arXiv:hep-th/0307274].
%
\bibitem{GTT}
 J.~P.~Gauntlett, D.~Tong and P.~K.~Townsend,
  Phys.\ Rev.\  D {\bf 63} (2001) 085001
  [arXiv:hep-th/0007124].
%
\bibitem{GPTT}
 J.~P.~Gauntlett, R.~Portugues, D.~Tong and P.~K.~Townsend,
  Phys.\ Rev.\  D {\bf 63} (2001) 085002
  [arXiv:hep-th/0008221].
%
\bibitem{ANNS}
 M.~Arai, M.~Naganuma, M.~Nitta and N.~Sakai,
  Nucl.\ Phys.\  B {\bf 652} (2003) 35
  [arXiv:hep-th/0211103].
%
\bibitem{AIN}
 M.~Arai, E.~Ivanov and J.~Niederle,
  Nucl.\ Phys.\  B {\bf 680} (2004) 23
  [arXiv:hep-th/0312037].
%
\bibitem{KG1}
  S.~J.~Gates,~Jr. and S.~M.~Kuzenko,
  Nucl.\ Phys.\ B {\bf 543} (1999) 122
  [arXiv:hep-th/9810137].
%
\bibitem{KG2}
  S.~J.~Gates,~Jr. and S.~M.~Kuzenko,
  Fortsch.\ Phys.\  {\bf 48} (2000) 115
  [arXiv:hep-th/9903013].
%
\bibitem{AKL1}
  M.~Arai, S.~M.~Kuzenko and U.~Lindstr\"om,
  JHEP {\bf 0702} (2007) 100
  [arXiv:hep-th/0612174].
%
\bibitem{AN}
  M.~Arai and M.~Nitta,
  Nucl.\ Phys.\ B {\bf 745} (2006) 208
  [arXiv:hep-th/0602277].
%
\bibitem{AKL2}
  M.~Arai, S.~M.~Kuzenko and U.~Lindstr\"om,
  JHEP {\bf 0712} (2007) 008
  [arXiv:0709.2633 [hep-th]].
%
\bibitem{KLR}
  A.~Karlhede, U.~Lindstr\"om and M.~Ro\v cek,
  Phys.\ Lett.\ B {\bf 147} (1984) 297.
%
\bibitem{LR}
  U.~Lindstr\"om and M.~Ro\v{c}ek,
  Commun.\ Math.\ Phys.\  {\bf 115} (1988) 21.
%
\bibitem{EINOOST}
  M.~Eto, Y.~Isozumi, M.~Nitta, K.~Ohashi, K.~Ohta, N.~Sakai and Y.~Tachikawa,
  Phys.\ Rev.\  D {\bf 71} (2005) 105009
  [arXiv:hep-th/0503033].
%
\bibitem{HN}
  K.~Higashijima and M.~Nitta,
  Prog.\ Theor.\ Phys.\  {\bf 103} (2000) 635
  [arXiv:hep-th/9911139].
%
\bibitem{SS}
  J.~Scherk and J.~H.~Schwarz,
  Phys.\ Lett.\  B {\bf 82} (1979) 60.
%
\bibitem{WB}
  J.~Wess and J.~Bagger,
  {\it Supersymmetry and supergravity},
{\rm  Princeton, USA: Univ. Pr. (1992).}
%
\bi{CV}
  E. Calabi and E. Vesentini,
  Ann. Math. {\bf 71}, 472 (1960).
%
\bi{KN} S. Kobayashi and K. Nomizu, {\it Foundations of
  Differential Geometry, Vol. II}, Wiley Interscience, New York (1996).
%
\bi{Hua1}  L. K. Hua,
 Ann. Math. {\bf 47} (1946) 167.
%
\bi{Hua2} L. K. Hua,
 {\it Harmonic Analysis of Functions of Several Complex Variables in the Classical Domains},
 American Mathematical Society, Providence (1963).
%
\bibitem{MPS}
  A.~Y.~Morozov, A.~M.~Perelomov and M.~A.~Shifman,
  Nucl.\ Phys.\ B {\bf 248} (1984) 279;
  A.~M.~Perelomov,
  Phys.\ Rept.\  {\bf 146} (1987) 135.
%
\bibitem{Kuzenko}
  S.~M.~Kuzenko,
  Phys.\ Lett.\  B {\bf 638} (2006) 288
  [arXiv:hep-th/0602050].
\end{thebibliography}
\end{document}